\documentclass[a4paper,11pt]{article}
\usepackage{subcaption}
\usepackage{jheppub} % for details on the use of the package, please see the JINST-author-manual
\usepackage{lineno}%\linenumbers
\usepackage{placeins}
\usepackage{amsmath}
\usepackage{style}
\usepackage{xspace}
\usepackage[normalem]{ulem}

\title{\boldmath Optimal-Transport-Based Cell Resampling for Negative and Pathological Event Weights}

\author{Regan~Doherty$^1$, Lauren~Hay$^{1,2}$, Rishabh~Jain$^1$, Matt~LeBlanc$^{1,2}$, Julia~Marrinan$^1$, Camille~Mauceri$^1$, Jennifer~Roloff$^1$}
\affiliation{$^1$Department of Physics, Brown University, Providence, RI, USA}
\affiliation{$^2$The NSF AI Institute for Artificial Intelligence \& Fundamental Interactions, Boston, MA, USA}

\emailAdd{rishabh\_jain@brown.edu, matt\_leblanc@brown.edu, jennifer\_roloff@brown.edu}

\abstract{
Negative and pathologically large Monte Carlo event weights strain the computing budgets of experiments at the Large Hadron Collider.
Cell resampling algorithms locally redistribute event weights among nearby events in a metric space.
We study the performance of metrics defined in terms of Optimal Transport, namely the Energy Mover's Distance and a spectral variant, in the context of such algorithms.
As these metrics are insensitive to the addition of soft and collinear radiation, they may be applied directly to particles at any stage of event generation.
When applied to samples simulated at next-to-leading order in quantum chromodynamics, this approach reduces the observed bias relative to other cell resampling techniques presented in the literature.
We also study the Cross-Section Mover's Distance as an unbinned, broadly applicable figure of merit for quantifying the bias introduced by any full-phase-space reweighting.
}

\begin{document}
\maketitle
\flushbottom

\clearpage

%%%%%%%%%%%%%%%%%%%%%%%%%%%%%%%%%%%%%%%%%%%%%%%%%%%%%%%%%%%%%%%%%%%%%%%%%%%%%%%
\section{Introduction}
\label{sec:intro}

Maximizing the scientific potential of the enormous datasets from the Large Hadron Collider (LHC)~\cite{Evans:2008zzb} requires significant computational resources for data acquisition, processing, and storage.
This need will intensify during the era of the High-Luminosity LHC (HL-LHC), when data collection rates will increase by roughly an order of magnitude relative to LHC Runs 2 and 3~\cite{Aberle:2749422}.
Nearly every analysis that is performed at the LHC is dependent on theoretical predictions, which are often implemented as Monte Carlo (MC) event generators.
The accuracy of these predictions, \emph{e.g.} when used for background estimation or calibration, often limits the precision of experimental results, and so realizing the highest levels of precision at the HL-LHC requires the adoption of the most sophisticated calculations that are available.
These simulations should be produced at scales large enough to have negligible statistical uncertainties relative to data, but their resultant footprint on-disk is already larger than that of the LHC data sample: by 2031, over 66\% of the estimated storage requirements for ATLAS during HL-LHC operations are expected to be used by simulated event samples~\cite{CERN-LHCC-2022-005}.
Both the CMS and ATLAS Phase-2 computing roadmaps indicate the risk of resource shortfall in terms of both computational and storage resources~\cite{CERN-LHCC-2022-005,Software:2815292}, motivating an aggressive R\&D campaign to reduce the demand on resources.

In leading-order (LO) MC generators, events are typically produced with uniform event weights.
The pursuit of higher perturbative accuracy at and beyond next-to-leading order ($\geq$NLO) necessitates the use of non-uniform event weights, which can become pathological when they are very large or negative.
Negative weights in particular degrade the statistical significance of an MC sample, since more events must be generated to achieve a target statistical uncertainty; the negative-weight fraction can reach $\sim30\%$ or more in large-scale samples produced by experimental collaborations~\cite{ATLAS:2021yza}.
Higher-order corrections account for quantum effects such as virtual particle loops and additional radiation, and their inclusion is essential for more precise theoretical calculations.
The NLO contributions from both virtual loop contributions and real emission of additional particles are individually infrared-divergent, but their divergences cancel in the sum.
To handle these individually divergent contributions in an MC framework, subtraction methods are employed: an approximate term is added and subtracted so that each piece can be integrated separately.
Widely-used subtraction procedures, such as Catani-Seymour dipole subtraction~\cite{Catani:1996vz} and the Frixione-Kunszt-Signer (FKS) method~\cite{Frixione:1995ms}, generate counter-events that carry negative weights.
These counter-events ensure that the inclusive cross section is correct, but can represent a non-trivial fraction of negatively-weighted events.

This situation is further complicated by the matching and merging procedures used to combine fixed-order matrix elements with parton shower algorithms.
These procedures are necessary because fixed-order calculations are accurate for well-separated, energetic emissions, while parton showers provide a good approximation for the soft and collinear radiation that dominates the structure of hadronic final states; combining the two requires care to avoid double-counting the emissions that both describe.
In widely-used MC\@NLO-type matching~\cite{Frixione:2002ik}, the parton shower approximation to the NLO real-emission contribution is subtracted to avoid double-counting of emissions.
This results in a negative weight whenever the shower overestimates the exact matrix element in a given region of phase space.
The FxFx~\cite{Frederix:2012ps} and CKKW-L~\cite{Lonnblad:2001iq,Lavesson:2005xu,Lonnblad:2011xx} multi-leg merging schemes can introduce additional negative weights through the application of scale vetoes and Sudakov form factors.

Events with weights that are negative and/or overly large are pathological in the context of data analysis and in the development of Artificial Intelligence and Machine Learning algorithms (AI/ML), where loss functions may not be easily applied to datasets with weights that are not strictly positive.
A sample with the negative weight fraction $f_{-}$ has its effective statistics reduced by a factor of $(1-2f_{-})^2$ relative to a sample of the same size with unit weights.
This compounding effect of more expensive $\geq$NLO calculations simultaneously requiring larger sample sizes presents a serious obstacle for computing related to the HL-LHC physics program.
One promising strategy to mitigate problems related to pathological weights is to locally redistribute event weights using a cell resampling algorithm, as proposed by Andersen \emph{et al.} in Refs.~\cite{Andersen:2021mvw,Andersen:2023cku,Andersen:2026ppe} and applied as an afterburner following sample generation.
These studies demonstrate that cell resampling can allow a smaller sample's statistical significance to match that of a much larger sample, even for complex multi-leg final states.
A critical component of this procedure is the distance metric used to define locality in phase space; the original implementations rely on a modified Euclidean metric that required clustering final-state particles into collimated groups known as \emph{jets} to ensure infrared and collinear (IRC) safety: the property that the metric is insensitive to the addition of soft particles and the splitting of one particle into two nearly-parallel ones.
In this paper, we propose performing cell resampling using metrics implemented with the particle physics application of the Earth Mover's Distance~\cite{10.1007/978-3-642-40020-9_43,192468,Rubner:1998:MDA:938978.939133,Rubner2000,10.1007/978-3-540-88690-7_37} -- or, Wasserstein distance~\cite{wasserstein1969markov,dobrushin1970prescribing} -- called the Energy Mover's Distance (EMD)~\cite{Komiske:2019fks}.
We also study a variant defined on event spectral functions known as the Spectral Energy Mover's Distance (sEMD)~\cite{Larkoski:2023qnv,Gambhir:2024ndc}.
These distances are computed by solving Optimal Transport (OT) problems, using standard implementations based on the Python Optimal Transport software library~\cite{flamary2017pot}.
Both the EMD and sEMD are IRC-safe metrics by design, and so we compare the performance of cell reweighting performed at different stages of event generation: directly following the hard-process matrix element calculation, after the parton shower, and after hadronization has occurred.
As a first step towards assessing the general applicability of such algorithms to a wide variety of final-state topologies, the performance of OT-based cell resampling is studied in final states including either $Z$ boson production in association with a jet, or top-antitop quark pairs.
Finally, to quantify the overall level of bias introduced by the cell reweighting procedure, we consider the Cross-Section Mover's Distance (\XMD) introduced in Ref.~\cite{Komiske:2020qhg} as a novel, holistic figure of merit.

\subsection{Cell resampling}

\begin{figure}
    \centering
    \includegraphics[width=0.9\textwidth]{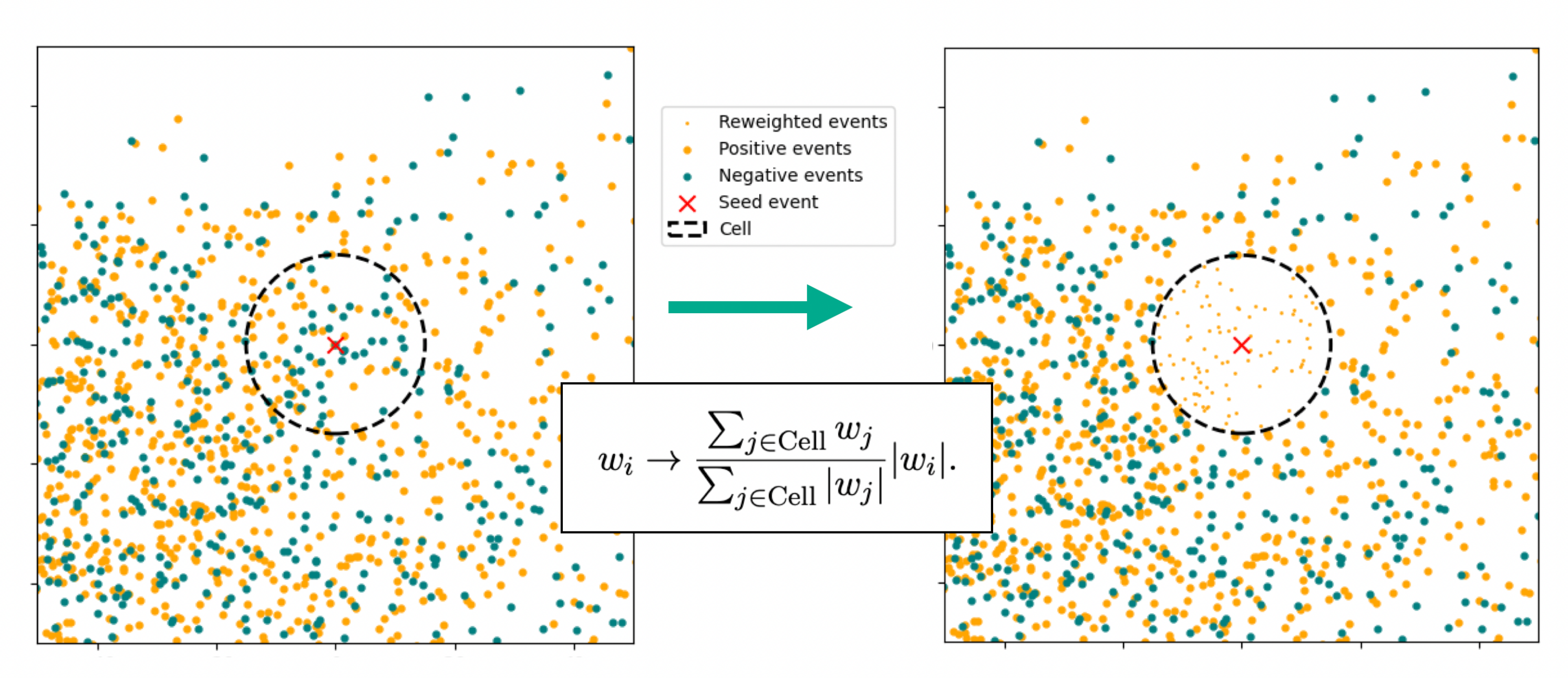}
    \caption{Simplified demonstration of cell-resampling in a \zjets sample with 37\% negative weights. The points are a multidimensional scaling of the events' pairwise EMD distances into a Cartesian space for illustration~\cite{Torgerson_1952}.}
    \label{fig:resampling}
\end{figure}

The goal of cell resampling is to modify event weights without adding bias to measurable observables.
This is done by redistributing the weights of negatively-weighted events with those of nearby events in phase space.
Consider a negatively-weighted event, referred to as the \emph{seed}, and a hypersphere of radius $R$ in phase space that is centered on it; this sphere defines the \emph{cell}.
As $R$ increases, more events will be contained in the cell.
When the total weight of events inside the cell is positive, the weights can be redistributed between events to make them all positive, while keeping the total weight of the cell constant. This is demonstrated in Figure~\ref{fig:resampling}.
One way to do this is with the following transformation: 
\begin{equation}
    w_i\rightarrow\frac{\sum_{j\in \mathrm{Cell}}w_j}{\sum_{j\in \mathrm{Cell}}|w_j|}|w_i|.
    \label{eq:crw}
\end{equation}
In the limit of a sufficiently large sample, the phase space density is high enough that successful reweighting can occur with vanishingly small values of $R$.
When $R$ becomes smaller than what is experimentally resolvable, the effect of this reweighting on physical observables becomes undetectable.

A critical aspect of the cell reweighting algorithm is the choice of distance metric between events, which determines which events are considered nearby and therefore reweighted together.
In the original proposal by Andersen \emph{et al.}~\cite{Andersen:2021mvw}, the distance was defined by summing modified Euclidean distances over different categories of final-state objects.
Consider the events $\mathcal{E}$ ($\mathcal{E'}$) with particle types $\{t_1^{(\prime)},t_2^{(\prime)},...\}$.
The distance $d$ takes the following form:
\begin{equation}
    d(\mathcal{E},\mathcal{E'}) = \sum_{i=1}^Td_t(t_i,t_i'). 
\end{equation}
If two sets of particles $t$ and $t'$ both contain $P$ particles with momenta $\{p_1^{(\prime)}, p_2^{(\prime)},...,p_P^{(\prime)}\}$, then the distance between them takes the following form:
\begin{equation}
    d_t(t,t') = \min_{\sigma\in S_P}\sum_{j=1}^P D(p_j,p'_{\sigma(j)}),
\end{equation}
where $S_P$ is the symmetric group, encoding the permutation invariance of sets, and $\sigma$ is a permutation.
$D$ is the modified Euclidean distance between particles given by
\begin{equation}
    D(p_j,p'_j)=\sqrt{|\vec{p}_j-\vec{p}_j'|^2+\tau^2(p_{T,j}-p'_{T,j})^2}.
\end{equation}
The parameter $\tau$ controls the relative weighting of angular and transverse-momentum differences in the metric; it was left at the default value of 0 for these studies.
If $t$ and $t'$ contain different numbers of particles, then a number of zero-momentum particles can be added until both sets have an equal number.
Without an initial clustering to reduce the number of particles, this distance metric is computationally expensive, as the number of permutations scales factorially with the number of particles.
In Ref.~\cite{Andersen:2023cku}, it was demonstrated that an approximate distance metric based on nearest neighbors can achieve $N\log(N)$ scaling through vantage point trees. 
A further refinement of the metric was introduced in Ref.~\cite{Andersen:2024mqh}, which resolves the particle types entering the metric calculation into hadronic objects (jets) and leptonic objects (dressed leptons), while also extending the set of infrared-safe objects to include isolated photons.
Since leptonic and hadronic final-state objects are assigned to distinct types, their contributions to the total distance $d(\mathcal{E},\mathcal{E}')$ are additive and separately controllable.
We adopt this leptonic/hadronic separation of the distance metric for our final results using this metric (Sec.~\ref{sec:results:SEMD}).

The principal limitation of this metric is that it is not IRC-safe: soft emissions and/or collinear splittings can substantially change the distance between two events even though these modifications are below the experimental resolution.
In Ref.~\cite{Andersen:2021mvw}, this was addressed by clustering final-state hadrons into IRC-safe jets prior to the reweighting procedure, and using these jets as the fundamental objects in the cell resampling algorithm.
However, this introduces a dependence on the jet definition, including parameters such as the jet radius and clustering algorithm~\cite{Salam:2010nqg} that directly affect the reweighting and therefore, all downstream observables.
The OT-based metrics used in this study, described in Section~\ref{sec:ot}, avoid this problem entirely: both the EMD and sEMD are IRC-safe by construction, allowing cell resampling to proceed directly on particles in an event without any intermediate clustering.
For these studies, with one notable exception discussed in Section~\ref{sec:results:betas}, the seeds are chosen randomly, avoiding an additional sorting step, consistent with the choice made in Ref.~\cite{Andersen:2021mvw}.

%%%%%%%%%%%%%%%%%%%%%%%%%%%%%%%%%%%%%%%%%%%%%%%%%%%%%%%%%%%%%%%%%%%%%%%%%%%%%%%
\section{Optimal-Transport-Based Metrics}\label{sec:ot}

This section describes the two optimal-transport based distance metrics used for cell resampling in this study.
Both metrics define an IRC-safe notion of distance between collider events by solving a transport problem: determining the minimum `work' required to rearrange the radiation pattern of one event into the other~\cite{Komiske:2019fks,Komiske:2020qhg}.
The EMD (Sec.~\ref{sec:emd}) operates directly on particles in the two-dimensional rapidity-azimuth plane, retaining full geometric information about the events.
The sEMD (Sec.~\ref{sec:semd}) instead operates on one-dimensional spectral representations of events, which encode the energy-weighted angular structure of particle pairs while automatically incorporating isometries such as rotations about the beam~\cite{Larkoski:2023qnv}.

\subsection{Energy Mover's Distance}\label{sec:emd}

The Energy Mover's Distance (EMD)~\cite{Komiske:2019fks,Komiske:2020qhg} is the particle physics equivalent of the Earth Mover's Distance~\cite{10.1007/978-3-642-40020-9_43,192468,Rubner:1998:MDA:938978.939133,Rubner2000,10.1007/978-3-540-88690-7_37}, which defines a metric between collider events by quantifying the minimum `work' needed to rearrange the radiation pattern of one event into another.
An event $\mathcal{E}$ with $N$ particles can be represented by its energy flow,
\begin{equation}
    \mathcal{E}(\hat{n})=\sum_{i=1}^{N}E_i\delta(\hat{n}-\hat{n}_i),
\end{equation}
where $E_i$ and $\hat{n}_i$ are the energy and direction of particle $i$, respectively, and the total energy is
\begin{equation}
    E_{\mathrm{tot}}=\int \mathcal{E}(\hat{n})d^2\hat{n}.
\end{equation}
The energy flow uniquely characterizes each event, and can be probed both experimentally and theoretically.
For proton-proton collisions at the LHC, where general-purpose detectors are cylindrical around the beam axis, particle positions are naturally parameterized in pseudorapidity ($\eta$) and azimuthal angle ($\phi$) plane, and the particle energy replaced by its transverse momentum \pt.

The EMD between two events, $\mathcal{E}$ and $\mathcal{E}'$, is defined as
\begin{equation}
    \mathrm{EMD}_{\beta,R}(\mathcal{E},\mathcal{E}')=\underset{[f_{ij}\geq0]}{\mathrm{min}}\sum_{i=1}^{N}\sum_{j=1}^{N'}f_{ij}\left(\frac{\theta_{ij}}{R}\right)^\beta + \left| \sum_{i=1}^NE_i -\sum_{j=1}^{N'}E_j'\right|,
    \label{eq:EMD}
\end{equation}
where $f_{ij}$ is the transport plan specifying how much energy is moved from particle $i$ of $\mathcal{E}$ to particle $j$ of $\mathcal{E}'$, subject to the constraints
\begin{equation}
    f_{ij}\geq0,\quad\sum_{j=1}^{N'}f_{ij}\leq E_i, \quad\sum_{i=1}^{N}f_{ij}\leq E_j', \quad\sum_{i=1}^{N}\sum_{j=1}^{N'}f_{ij}=\mathrm{min}(E_\mathrm{tot},E'_\mathrm{tot}).
\end{equation}
The first term in Eq.~\ref{eq:EMD} quantifies the cost of rearranging the radiation pattern of one event into the other, and the second term accounts for any difference in the total energy between events.
Three user-defined inputs specify the metric.
The first is the ground metric $\theta_{ij}$, which defines the pairwise distances between particles.
We choose this to be the Euclidean distance in the pseudorapidity-azimuth plane, $$\theta_{ij}=\sqrt{(\phi_i-\phi'_j)^2+(\eta_i-\eta'_j)^2},$$ though more generally any function satisfying the metric axioms is a valid choice~\cite{Komiske:2019fks}.
The second parameter, $R>0$, sets the relative importance of the transport and total-energy terms in Eq.~\ref{eq:EMD}.
Finally, the angular exponent $\beta>0$ controls the sensitivity of the metric to transport at different angular scales.
Larger values of $\beta>0$ penalize large-angle transport more heavily while making small-angle rearrangements essentially free; smaller values of $\beta$ weight distances more uniformly across angular scales~\cite{ATLAS:2023mny,Cesarotti:2020xtf}.
When $\beta>1$, Eq.~\ref{eq:EMD} must be raised to the power $1/\beta$ to satisfy the generalized triangle inequality and constitute a true metric, and $R$ must be at least half of the maximum possible ground space distance.

An exact EMD computation scales as $\mathcal{O}(N^3\log(N))$ for events with $N$ particles, though approximate solvers can reduce the computational complexity to $\mathcal{O}(N^2)$.
While this complexity is still challenging for large particle multiplicities, further improvements related to runtime efficiency have been left to future studies.
We have used an implementation of the EMD based on the Python Optimal Transport library~\cite{flamary2017pot}.
In this study, we set $R$ equal to the maximum possible ground space distance given the pseudorapidity acceptance of the ATLAS and CMS detectors at the LHC ($R=11.64$), and use $\beta=1$ as the baseline metric.
This makes the EMD equivalent to the first Wasserstein distance~\cite{wasserstein1969markov,dobrushin1970prescribing}.
The impact of changing the value of $\beta$ on cell resampling performance is studied in Sec.~\ref{sec:results:betas}.

\subsection{Spectral Energy Mover's Distance}\label{sec:semd}

The Spectral Energy Mover's Distance (sEMD), introduced in Ref.~\cite{Larkoski:2023qnv}, provides an alternative OT-based metric between collider events that operates on a one-dimensional representation rather than in the full two-dimensional pseudorapidity--azimuth plane.
This lower-dimensional formulation retains IRC-safety and leads to significant computational advantages, particularly when implemented through the \textsc{Specter} framework~\cite{Gambhir:2024ndc}.

The key idea is to represent each event not by the positions and energies of individual particles, but by its \emph{spectral function}: a one-dimensional distribution that encodes the energy-weighted pairwise angular structure of the event.
For an event $\mathcal{E}$ with $N$ particles, each pair $(i,j)$ of particles defines a pairwise angular distance $\omega_{ij}$ and carries an energy weight proportional to the product of particle energies $E_iE_j$.
The spectral function is then defined as
\begin{equation}
 s(\omega) = \sum_{i,j\in\mathcal{E}} E_i E_j\,
    \delta(\omega - \omega_{ij}),
    \label{eq:spectral}    
\end{equation}%
Because $s(\omega)$ depends only on pairwise angular distances and is multi-linear in particle energies, it is manifestly IRC-safe.
Furthermore, it is invariant under isometries of the detector geometry, such as rotations about the beam, as it is constructed using only relative angular distances.

The spectral function's cumulative distribution is given by
\begin{equation}
S(\omega) = \int_0^{\omega} d\omega'\, s(\omega') 
  = \sum_{i,j} E_i E_j\,\Theta(\omega - \omega_{ij}),    \label{eq:spectral:cumulative}    
\end{equation}
and it is normalized to the square of the total energy,
\begin{equation}
 S(\omega_{max}) = E_{tot}^2,
    \label{eq:spectral:norm}    
\end{equation}
which provides a monotonically increasing function that can be used to define optimal transport in one dimension.
The inverse cumulative distribution, $S^{-1}(E^2)$, therefore provides the value of $\omega$ enclosing a given squared energy $E^{2}$.

The sEMD between two events $\mathcal{E}$ and $\mathcal{E'}$ can be defined as the transport cost between two of these inverse cumulative distributions:
\begin{equation}
 \mathrm{sEMD}(\mathcal{E},\mathcal{E'})_{p} = \int_0^{E^2_{tot}}dE^2|S^{-1}_{\mathcal{E}}(E^2)-S^{-1}_\mathcal{E'}(E^2)|^p.
    \label{eq:spectral:semd:p}    
\end{equation}
A key result of Ref.~\cite{Gambhir:2024ndc} is that this integral admits a closed-form evaluation for events represented by discrete sets of particles, enabling evaluation of the sEMD with $p=2$ in $\mathcal{O}(N^2\log(N))$ runtime.

The sEMD shares important properties with the EMD (Sec.~\ref{sec:emd}): both are IRC-safe, both scale bilinearly with particle energy, and both have been shown to produce highly correlated distance measurements between jets of particles~\cite{Gambhir:2024ndc}.
However, the two metrics do introduce subtly different topologies on the space of events.
Certain configurations that are well-separated in the full two-dimensional energy-flow space can appear nearby in spectral space, and \emph{vice-versa}~\cite{Larkoski:2023qnv}.
These differences are most pronounced for events with hard, wide-angle radiation and are suppressed in strongly-ordered configurations, \emph{e.g.} those characteristic of jet substructure~\cite{Kogler:2018hem}.
The practical impact of these differences in terms of cell resampling algorithms is discussed later in this work.

%%%%%%%%%%%%%%%%%%%%%%%%%%%%%%%%%%%%%%%%%%%%%%%%%%%%%%%%%%%%%%%%%%%%%%%%%%%%%%%

\section{Simulated event samples}\label{sec:mc}

The studies in this paper use simulated proton--proton collision events\footnote{For readers interested in an overview of MC simulation in high-energy physics, a recent review may be found in Ref.~\cite{vanBeekveld:2026uxl}.} at a center-of-mass energy of $\sqrt{s}=13$~TeV, generated for two distinct final states: the production of a $Z$ boson in association with jets (\zjets) and top quark pair (\ttbar) production.
These two processes represent qualitatively different final-state topologies: \zjets events are dominated by QCD radiation recoiling against a color singlet, while \ttbar events produce a more complex energy flow with higher particle multiplicities and additional hard scales from the top quark decays.
Together, they are used to provide an initial assessment of whether OT-based cell resampling generalizes across different event topologies.

For both processes, hard-scattering matrix elements are calculated at next-to-leading order (NLO) QCD using \textsc{MadGraph5\_aMC@NLO}~3.5.6~\cite{Alwall:2014hca}.
The NNPDF2.3 NLO parton distribution function set~\cite{Ball:2012cx} is used.
Renormalization and factorization scales are set dynamically to half the sum of the transverse masses of the final-state particles, following the \textsc{MadGraph5\_aMC@NLO} default.
The NLO matrix elements are matched to the parton shower using the MC@NLO method~\cite{Frixione:2002ik}; no multi-leg merging is applied.

Parton showering, hadronization, multi-parton interactions and the underlying event are modeled with \textsc{Pythia}~8.3~\cite{Bierlich:2022pfr} using the default Monash tune~\cite{Skands:2014pea}.
Hadronization is performed using \textsc{Pythia}'s default Lund string fragmentation model~\cite{Andersson:1983ia,Sjostrand:1984ic}.
No folding~\cite{Frederix:2020trv} of the NLO subtraction terms is applied, so the samples retain their full negative weight fraction as produced by the MC@NLO matching procedure.

For both processes, events are stored at three successive stages of the simulation: immediately after the hard-scattering matrix element calculation (`HS'), after the parton shower (`PS'), and after hadronization (`HAD').
This enables the comparison of cell resampling performance at different simulation stages presented in Sec.~\ref{sec:results:stages}, and is a key advantage of using inherently IRC-safe distance metrics that can be applied directly to particle-level information at any stage.
The parton shower and hadronization algorithms do not introduce any additional negatively weighted events in these samples, and so comparisons are consistently made using the same event samples.

For both \zjets and \ttbar production, $10^5$ events were generated.
Both leptonic and hadronic decays of the $Z$ boson and top quark were allowed.
A loose selection is applied at the generator level to ensure that event generation is efficient but still relatively inclusive: jets are clustered using the $k_t$ algorithm~\cite{Catani:1993hr} with a radius parameter of $R=0.7$ and are required to have \pt $> 10$~GeV; same-flavor opposite-sign lepton pairs are required to have an invariant mass $m_{\ell\ell} > 30$~GeV.
At the analysis level, final-state particles are required to have \pt$> 0.1$~GeV and $|\eta| < 4.9$.
Jets are reconstructed using the anti-$k_t$ algorithm~\cite{Cacciari:2008gp} with a radius parameter of $R=0.4$, as implemented with the \textsc{FastJet} software package~\cite{Cacciari:2011ma}.
Selected jets are required to have \pt$> 20$~GeV and to be within $|\eta|<4.5$.
Following the complete event generation procedure, the resulting \zjets and \ttbar samples have negative weight fractions of $37.6\%$ and $22.8\%$, respectively.

%%%%%%%%%%%%%%%%%%%%%%%%%%%%%%%%%%%%%%%%%%%%%%%%%%%%%%%%%%%%%%%%%%%%%%%%%%%%%%%
\section{Figures of Merit}\label{sec:fom}

The performance of the cell resampling algorithm with different metric choices is primarily assessed in terms of two figures of merit.
First, the impact of reweightings on individual kinematic distributions is evaluated by comparing the reweighted samples to the original ones across a set of benchmark kinematic distributions, which differ depending on the final state being studied (Sec.~\ref{sec:fom:observables}).
Second, the `Cross-Section Mover's Distance' is introduced as a new, holistic figure of merit to quantify the overall level of distortion of the reweighted sample in a way that does not depend on a predetermined choice of observables or binning (Sec.~\ref{sec:fom:xmd}).

As described in Sec.~\ref{sec:mc}, simulated event samples of both \zjets and \ttbar events are used to assess the performance of the cell resampling algorithms across different topologies.

\subsection{Kinematic Distributions}\label{sec:fom:observables}
The effect of cell resampling on the modeling of kinematic distributions is evaluated using a set of observables that probe different physical effects.
Two observables are common to the \zjets and \ttbar final states, and additional process-specific observables are included to test the sensitivity of features that have particular physics relevance.

The scalar sum of the transverse momentum of all selected jets, \Ht, is sensitive to the overall energy scale of the event but relatively insensitive to the angular distribution of energy within them.
The jet multiplicity, \njets, is IRC-sensitive to threshold effects from the jet \pt $>20$~GeV requirement; this sensitivity makes it a useful stress-test of the cell reweighting procedure.

For \zjets, the angular distance between the two leading jets $j_1$ and $j_2$ in the event \drjj and their \pt ratio \ptrat respectively probe the angular structure and energy sharing of the hardest hadronic emissions in the event.

For \ttbar events, the minimum invariant mass constructed of a $b$-jet and lepton \mbl is also studied in place of the two observables specific to \zjets.
This observable has a kinematic endpoint at $\sqrt{m_t^2-m_W^2} \approx 153$~GeV, and its shape is directly sensitive to $m_t$; it has been used in precision top mass ($m_t$) measurements by both CMS and ATLAS~\cite{CMS:2017znf,ATLAS:2016muw}.
Any bias introduced by cell resampling in the \mbl distribution could thus propagate into future measurements that apply this technique.

\subsection{Cross-section mover's distance}\label{sec:fom:xmd}

The Cross-Section Mover's Distance (\XMD) was first proposed in Ref.~\cite{Komiske:2020qhg} as a way to define a metric on the space of quantum field theories; here, we introduce it as a practical tool to quantify the distance between two unbinned samples of events in terms of an OT problem at the aggregate sample level.

Obtaining a reliable quantification of the full difference between two samples of events is challenging. 
While many metrics have been proposed, none of these fully satisfy the needs of this study.
Integral probability metrics~\cite{Muller_1997} and $f$-divergences do not scale easily to the low-level, high-dimensional data that exists in collider events.
Commonly used $f$-divergences, such as Kullback-Leibler~\cite{10.1214/aoms/1177729694}, Jensen-Shannon~\cite{61115,4767707}, and the Pearson $\chi^2$~\cite{Pearson01071900} rely on choosing individual observables to calculate these divergences.
Classifier-based methods typically do not account for the presence of negatively weighted events~\cite{Krause:2021ilc}, though some recent estimators can account for this~\cite{Drnevich:2024vfj}. 
Even when they correctly include weighted events, these results can be difficult to reproduce, hard to interpret, and may not be sensitive to certain types of effects~\cite{Kansal:2022spb}. 

The \XMD between two samples is defined analogously to the EMD between two events: just as the EMD quantifies the minimum cost of rearranging the energy flow of one event to match another, the \XMD quantifies the minimum cost of rearranging the cross-section distribution of one sample to match another.
Concretely, consider two samples of events, $\{(\mathcal{E}_i, \sigma_i)\}$ and $\{(\mathcal{E}'_j, \sigma'_j)\}$, where $\sigma_i$ is the cross-section weight associated with event $\mathcal{E}_i$.
The \XMD takes the following form, where event-level analogues to quantities defined in Sec.~\ref{sec:emd} are denoted by capital letters:
\begin{equation}
    \text{\XMD}_{\gamma,S} = \min_{[F_{ij}\geq 0]}
    \sum_{i=1}^{N}\sum_{j=1}^{M}
    F_{ij} \frac{\Theta_{ij}^\gamma}{S^\gamma}\,
    + \left|\sum_{i=1}^{N}\sigma_i
    - \sum_{j=1}^{M}\sigma'_j\right|.
    \label{eq:xmd}
\end{equation}
Here, $F_{ij}$ is the \emph{event-level} transport plan specifying how much cross section is moved from event $i$ in one sample to event $j$ in the other, subject to constraints analogous to those in the EMD (Eq.~\ref{eq:EMD}).
The ground metric $\Theta_{ij}$ is an \emph{event-level} EMD,
\begin{equation}
    \Theta_{ij} = \mathrm{EMD}_{\beta,R}
    (\mathcal{E}_i, \mathcal{E}'_j),
    \label{eq:xmd:ground}
\end{equation}
so that the cost of transporting cross section between two events is determined by how distant those events are in the ground space.
The parameters $\gamma$ and $S$ play roles analogous to $\beta$ and $R$ in the event-level EMD: $\gamma$ controls the sensitivity to transport in the ground space and $S$ sets the relative importance of the transport and total-cross-section terms.
In these studies, the \XMD $\gamma$ parameter is set to 1 and $S$ is set to the maximum distance in the EMD distance matrix, normalizing the range of \XMD values to be between 0 and 1.
The \XMD is zero when two samples are identical and grows as they become more different, providing an inclusive measure of sample-level distortion that is sensitive to all kinematic features simultaneously rather than to any particular projection.
Applied to cell resampling, a smaller \XMD between the original and reweighted samples indicates that less overall bias has been introduced.
A small \XMD does not preclude the presence of a significant bias in a given marginalized distribution: the observable-level comparisons of Sec.~\ref{sec:fom:observables} remain essential, and all figures of merit should be considered together when assessing the performance of a given reweighting.

\subsubsection{Treatment of negative event weights with the \XMD}

A practical complication in evaluating the \XMD arises from the presence of negatively-weighted events in the original sample.
The standard OT formulation assumes all weights are non-negative, and so the transport plan $F_{ij}$ and marginal constraints in Eq.~\ref{eq:xmd} are invalid when events carry negative cross section.
To handle this, we use a simple and efficient approach: a constant offset $c$ is added to every event weight in both samples before computing the \XMD.
Specifically, we choose $c=|\min_i{\sigma_i}|$ for the set of $i$ events, so that all weights in the original sample become non-negative ($\sigma_i+c\geq0\,\forall\,i$).
The same offset is added to every event in the reweighted sample, and the \XMD is then computed using standard OT on the resulting pair of non-negative distributions:
\begin{equation}
    \text{\XMD}\left(\{(\mathcal{E}_i, \sigma_i)\},\, \{(\mathcal{E}'_j, \sigma'_j)\}\right)
    = \text{\XMD}\left(\{(\mathcal{E}_i, \sigma_i + c)\},\, \{(\mathcal{E}'_j, \sigma'_j + c)\}\right).
    \label{eq:xmd:offset}
\end{equation}
Adding the same offset to every event in both samples leaves the transport cost unchanged: the extra weight enters both sides of the transport problem identically and cancels.
Formally, this is a cancellation property of the generalized Wasserstein distance $W$ that was proven in Refs.~\cite{2019arXiv191005105P,Piccoli:2016properties}: for any measures $\mu$, $\nu$ and $\eta$, the transport cost satisfies $W(\mu+\eta,\nu+\eta)=W(\mu,\nu)$.
The offset $c$ plays the role of $\eta$, so the equivalence holds only when both samples share the same support.
In the case of cell resampling, which reweights events without adding or removing them, this assumption is valid.
In the case of samples with differing support, ghost events could be included to apply the offset over a common set of events.

This constant-offset procedure has two practical advantages over a decomposition into positive and negative subsets.
First, it avoids the need to explicitly separate the samples into positively- and negatively-weighted subsets and recombine them, reducing bookkeeping~\cite{Drnevich:2025evh}.
Second, it allows the standard (positive-measure) OT solver to be used without modification, since all weights passed to the solver are non-negative by construction.

\section{Results}\label{sec:results}

In this section, a comparison of different cell reweighting configurations will be made using the Monte Carlo generated event samples described in Sec.~\ref{sec:mc}, according to the figures of merit outlined in Sec.~\ref{sec:fom}.
Different metric choices will result in inter-event distances with different scales.
Therefore, when comparisons between metrics are made, they are done for a given fixed reweighting fraction \frw to ensure that such comparisons are justified.

\subsection{Reweighting at different event generation stages}\label{sec:results:stages}

Previous cell-reweighting studies have been performed exclusively on fully hadronized events to ensure that the procedure is IRC-safe~\cite{Andersen:2021mvw}.
As the OT-based metrics we consider are themselves IRC-safe, they can be directly applied to events after any stage of event generation without intermediate jet clustering.
We compare the performance of the reweighting at three stages of event generation: at Born level after the hard-scattering matrix element calculation (HS), after the parton shower (PS), and after hadronization (HAD).
Each subsequent stage of simulation adds complexity and information to the event, but also increases the particle multiplicity.
Comparing reweighting performed at different stages of event generation assesses the trade-off between the use of a more physical event representation and the computational complexity of the OT calculation.
While the reweighting is performed after different stages of event generation, the performance is always assessed on a fully simulated event (\emph{i.e.} post-hadronization).
In this comparison, the angular exponent $\beta$ of the EMD calculation is set to 1.0; the effect of varying this parameter is studied in Sec.~\ref{sec:results:betas}.

\begin{figure}[htbp]
\centering
   \includegraphics[width=0.67\textwidth]{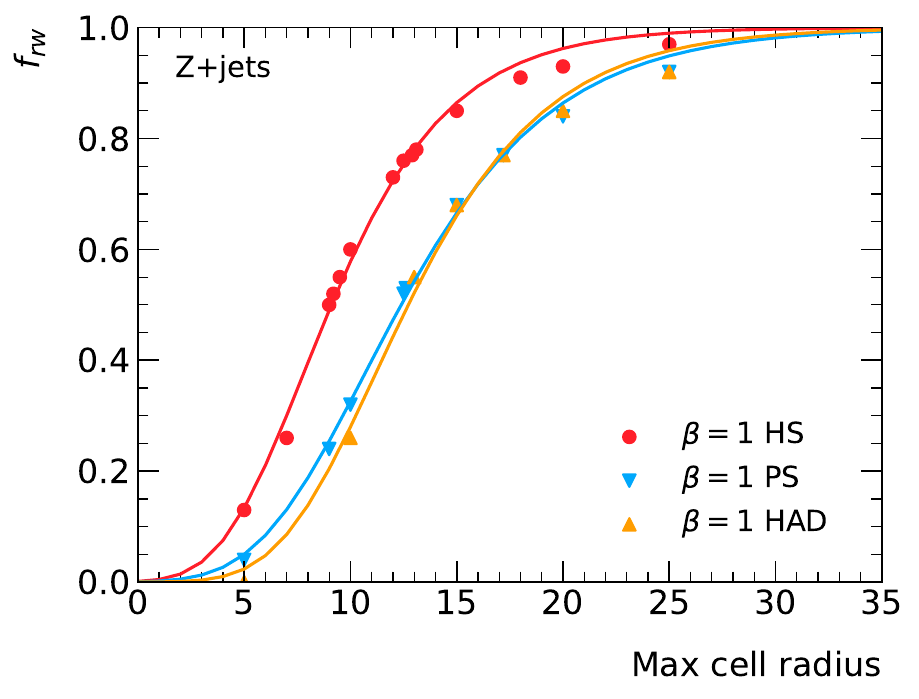}
   \caption{The distribution of \frw as a function of the maximum cell radius for \zjets events reweighted at matrix-element, parton shower, and hadronization level.
The distributions are fit using a Richards' curve~\cite{10.1093/jxb/10.2.290}.}
 \label{fig:stages:reweightfrac}
\end{figure}

Figure~\ref{fig:stages:reweightfrac} shows the monotonically-increasing fraction of negatively-weighted events that have been reweighted to have a positive value as a function of the cell radius, for each of the three reweighting options.
As events are reweighted, the average absolute value of the weights decreases, which means that typically more events must be included within a cell to produce a positive average weight.
The parton shower and hadronization reweightings show similar dependence on the cell radius, while reweighting immediately after the hard process can be performed with a smaller radius to achieve the same \frw.

\begin{figure}[htbp]
\centering
     \subfloat[\label{fig:stages:cellevents}]{
   \includegraphics[width=0.45\textwidth]{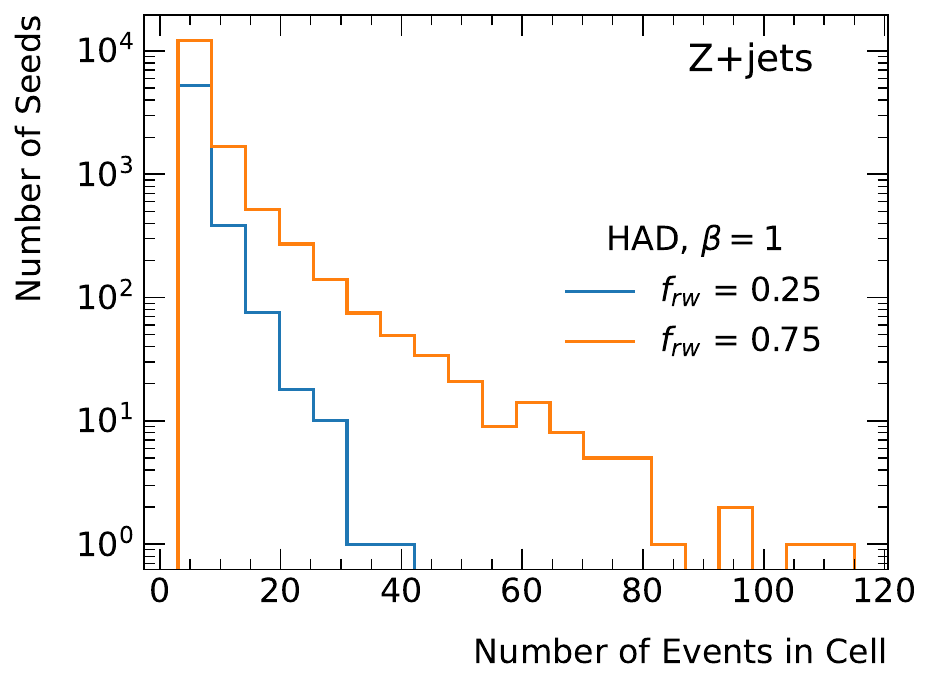}
    }     
    \subfloat[\label{fig:stages:cellradii}]{
   \includegraphics[width=0.45\textwidth]{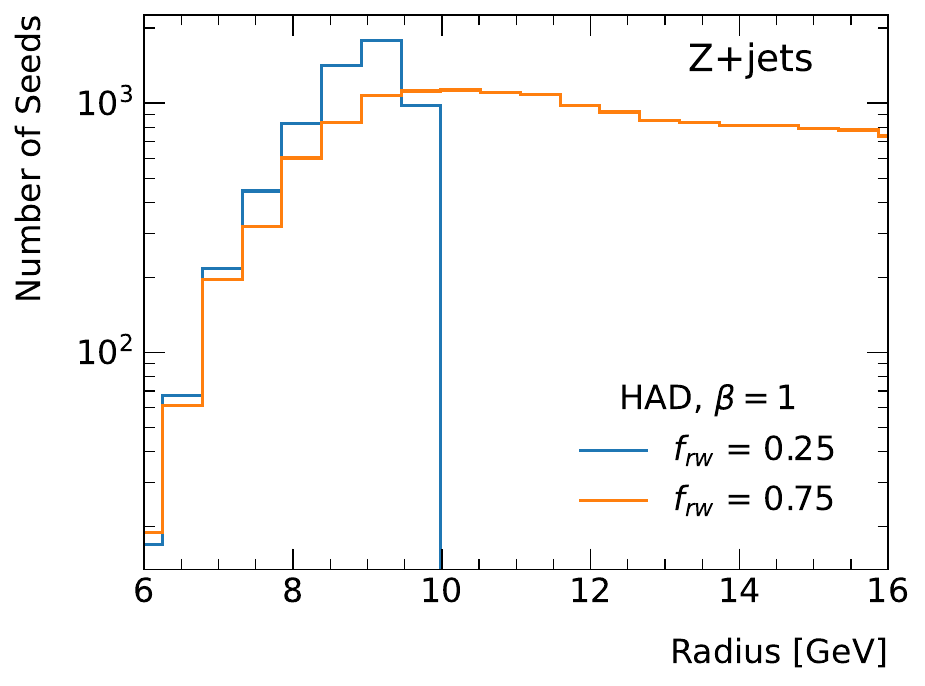}
    }
   \caption{(\ref{fig:stages:cellevents}) The number of events included in a reweighting for a given seed, and (\ref{fig:stages:cellradii}) the distribution of the cell radii for \frw$=0.25$ and \frw$=0.75$, using EMDs with $\beta=1$ and hadronized information to perform the reweighting.}
 \label{fig:stages:cell}
\end{figure}

Figure~\ref{fig:stages:cell} shows the number of events in a reweighting cell and the distribution of cell radii for a reweighting performed based on the HAD event for two fixed reweighting fractions, \frw$=0.25$ and \frw$=0.75$.
For these samples, the initial event weights have the same magnitude and are either negative or positive.
This means that the first set of reweighted events will have cells with fewer events in them, as the cell radius only needs to contain two positively weighted events; this is observed in the cell radius distribution, where the \frw$=0.75$ reweighting peaks around 8 GeV instead of 6 GeV.
As the reweighting continues and a larger fraction of events have numerically smaller weights, the number of events per cell required for positivity increases, reaching up to $\sim100$ events per cell in some cases.
Since the seed for the reweighting is chosen randomly, the differences between these distributions stem from the changes to nearby event weights, not from an inherent bias in the underlying distributions.

\begin{figure}[htbp]
\centering{
\includegraphics[width=0.67\textwidth]{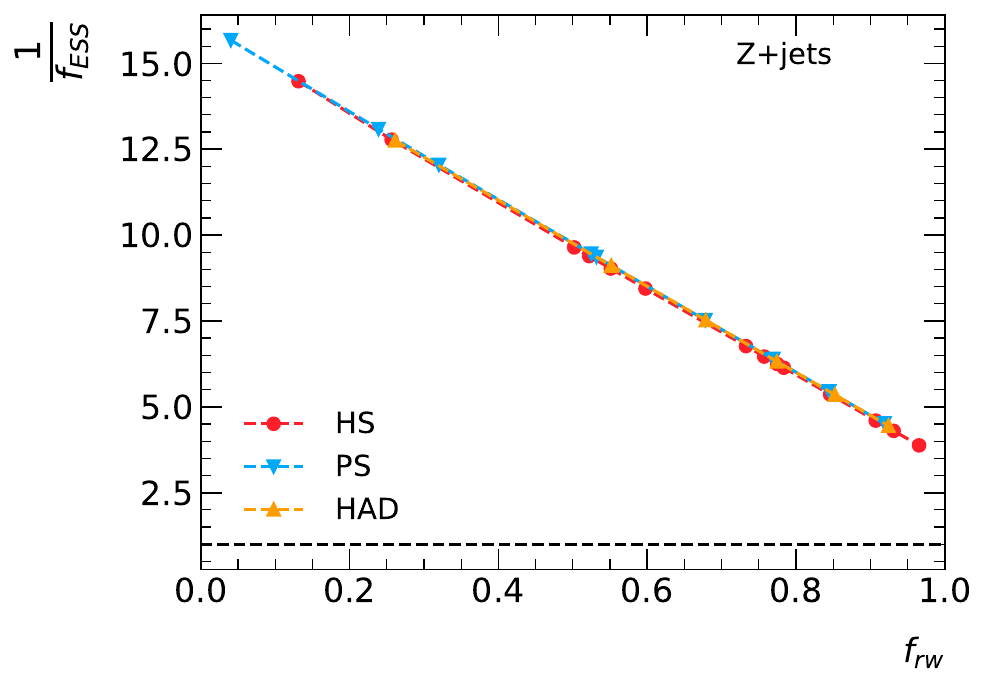}
   \caption{The value of $1/$\fess as a function of \frw for events reweighted using HS, PS, and HAD events. The dashed black line shows the value $1/$\fess for a uniformly weighted sample.}
    \label{fig:dilution}
    }
\end{figure}

When a fraction $\varepsilon$ of generated events carry negative weights, positive and negative contributions partially cancel in each bin, reducing the effective statistical power of the sample.
The Kish effective sample fraction \fess~\cite{bimj.19680100122} proposed in Ref.~\cite{Farkh:2026mtw} quantifies the relative statistical power of a sample compared to that of a fully unweighted sample with the same number of events:
\begin{equation}
    f_{ESS} = \frac{1}{N} \frac{(\sum_i w_i)^2}{\sum_i w_i^2},
\end{equation}
where $w_i$ is the weight of event $i$ and $N$ is the total number of events.
For the \textsc{MadGraph5\_aMC@NLO} samples used in this study (Sec.~\ref{sec:mc}), all initial event weights have the same magnitude, and so this simplifies to be 
\begin{equation}
    f_{ESS}(\varepsilon) = (1-2\varepsilon)^2,
\end{equation}
 where $\varepsilon$ is the fraction of negatively weighted events~\cite{Frederix:2020trv,Danziger:2021xvr}.
The \zjets sample used in these studies has a negative-weight fraction of $\varepsilon = 37.6\%$, so generating an equivalent unweighted sample requires $1/f_{ESS}(\varepsilon) = 16$ times as many events.
Figure~\ref{fig:dilution} shows $1/$\fess as a function of \frw.
With \frw$=0.25$ or $0.75$, $1/$\fess is respectively reduced from 16 to 12 or 6, resulting in a meaningful improvement in the statistical power of the samples.

\begin{figure}[htbp]
\centering{
   \subfloat[]{\includegraphics[width=0.45\textwidth]{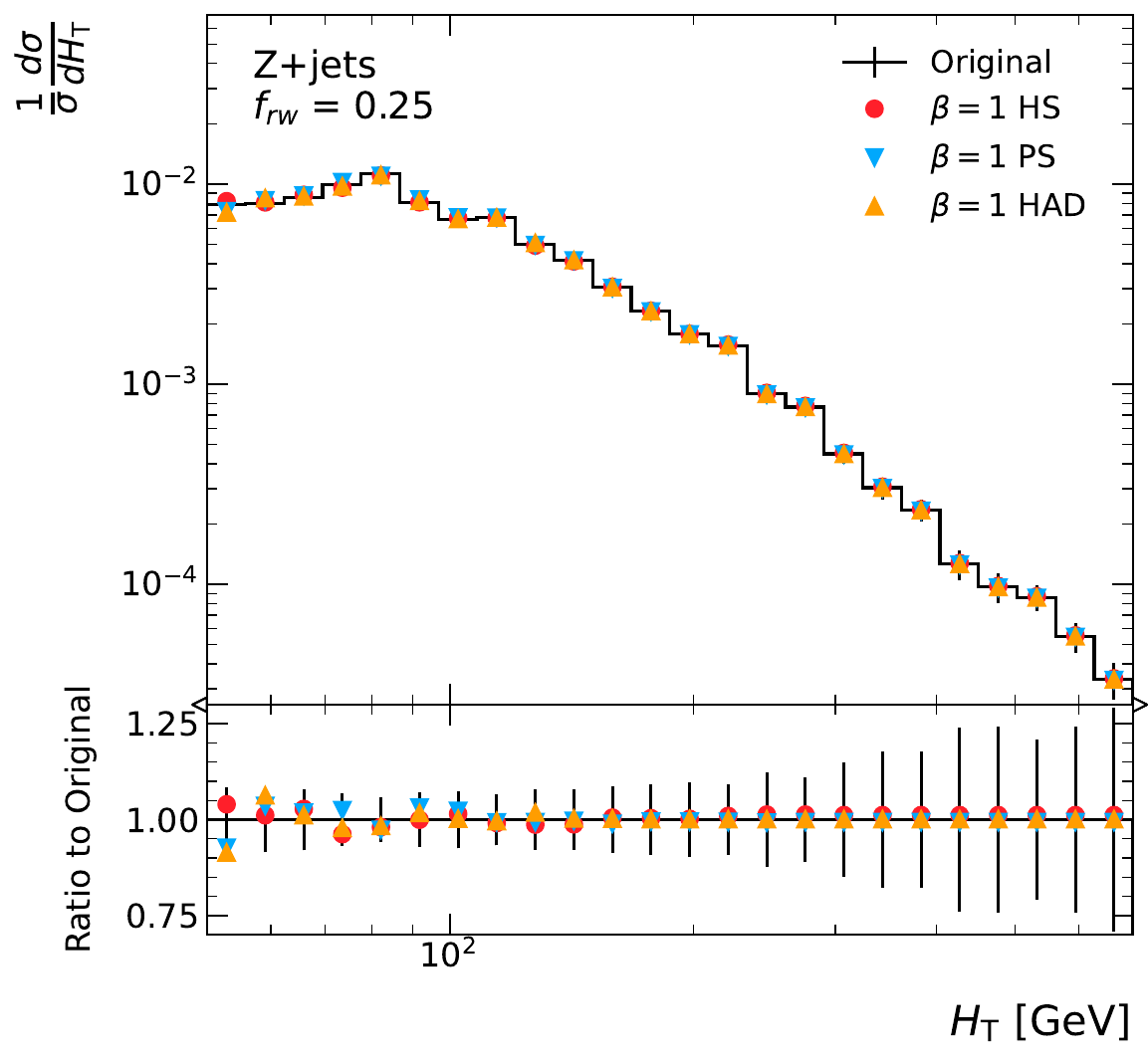}\label{fig:stages:ht:25}}
    \subfloat[]{\includegraphics[width=0.45\textwidth]{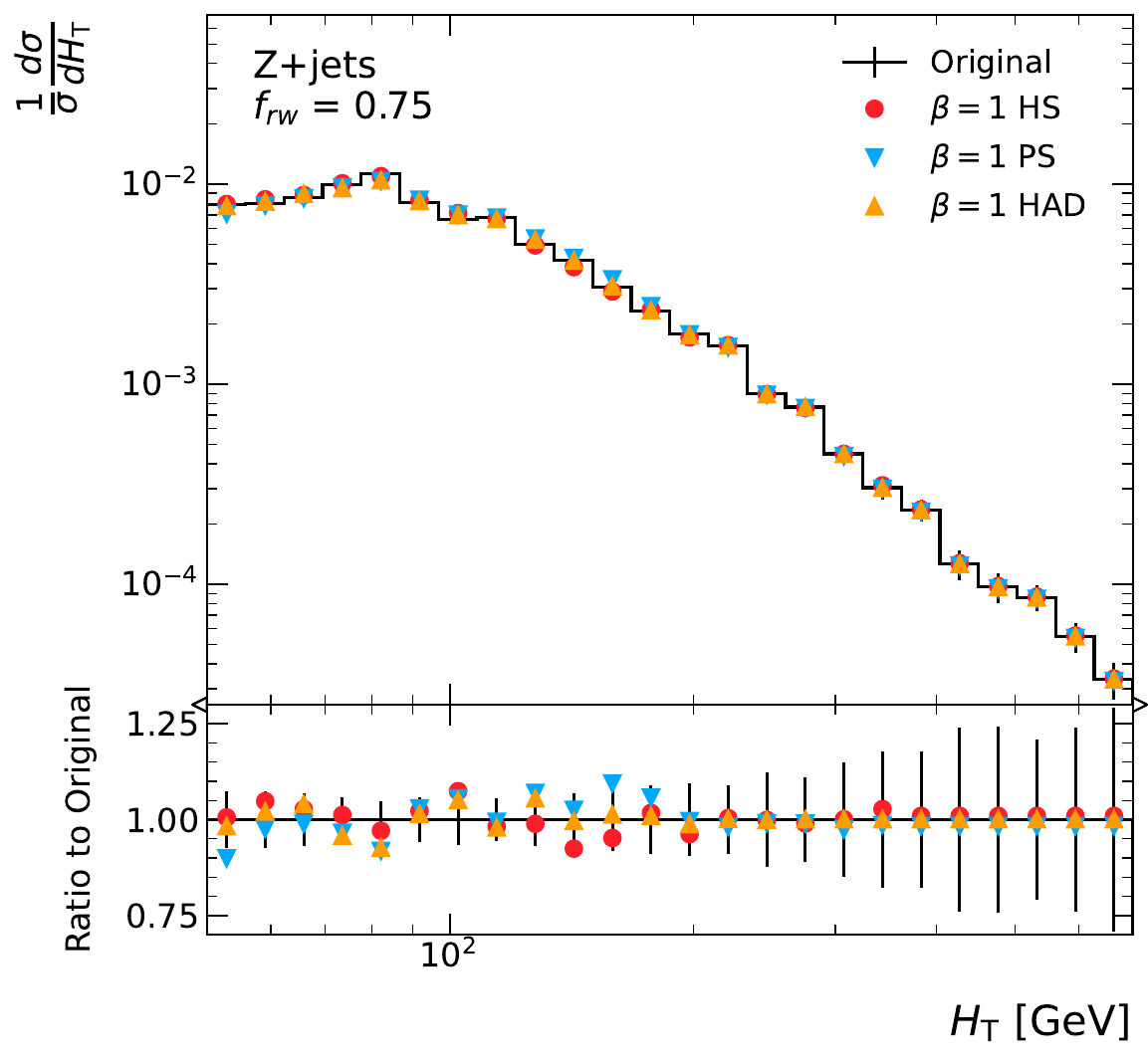}\label{fig:stages:ht:75}} \\
   \subfloat[]{\includegraphics[width=0.45\textwidth]{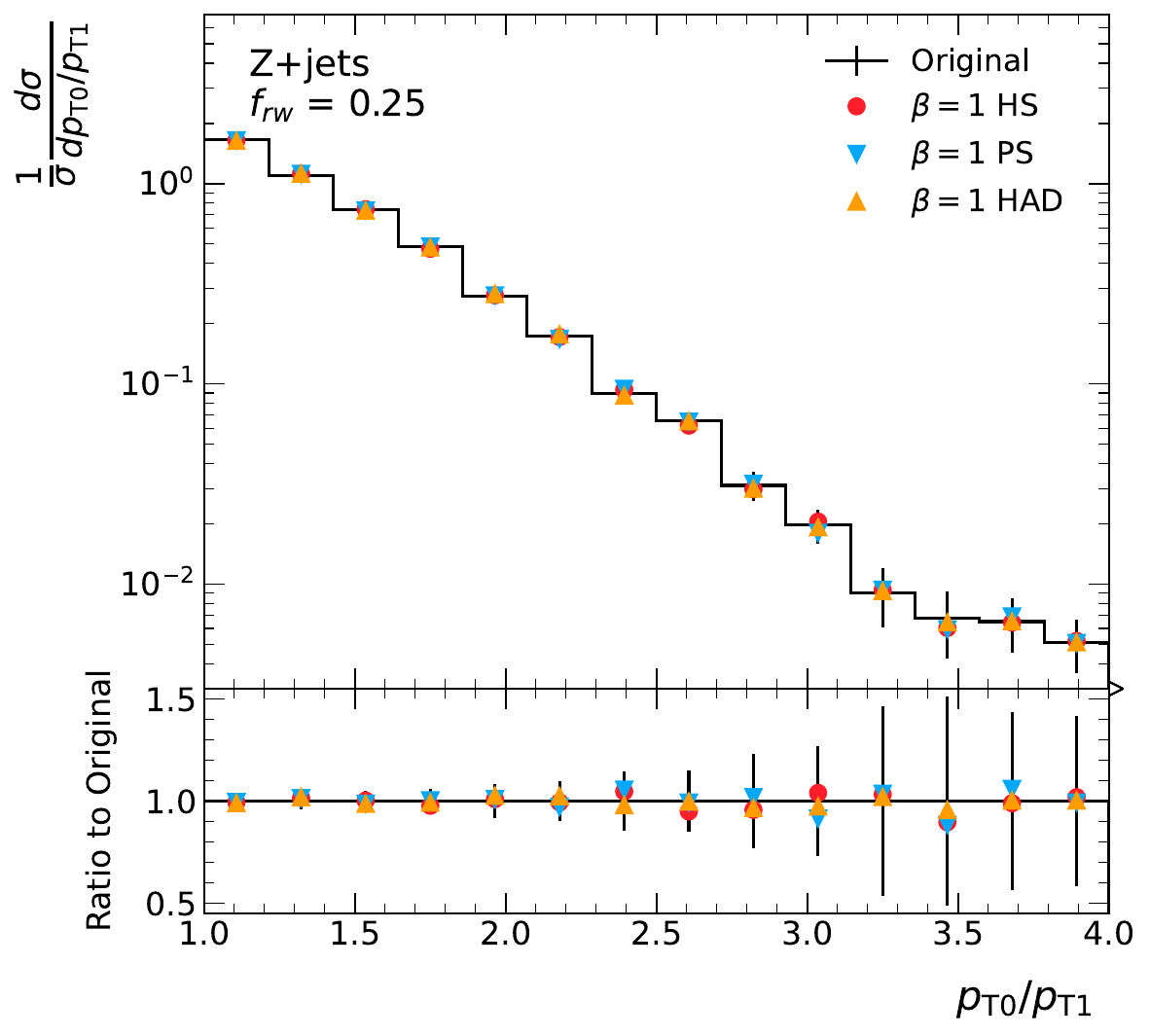}\label{fig:stages:ptrat:25}}
    \subfloat[]{\includegraphics[width=0.45\textwidth]{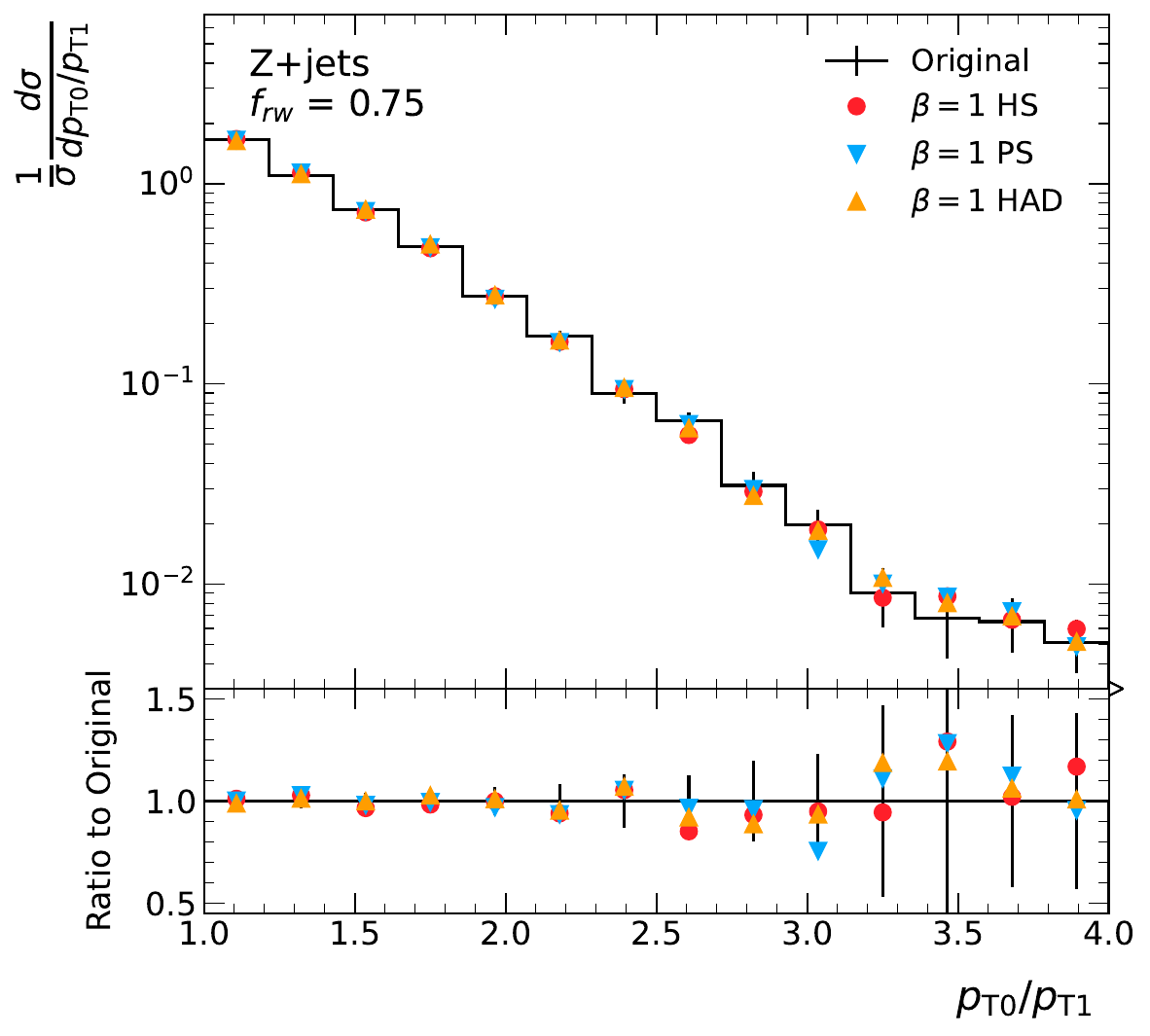}\label{fig:stages:ptrat:75}}
   \caption{Comparison of hadron-level unweighted \zjets  (\ref{fig:stages:ht:25}, ~\ref{fig:stages:ht:75}) \Ht and (\ref{fig:stages:ptrat:25}, \ref{fig:stages:ptrat:75}) \ptrat distributions to samples with \frw$=0.25$ (left) and \frw$=0.75$ (right) using HS, PS, and HAD event weights with $\beta=1$.
}
 \label{fig:stages:beta1:1}
 }
\end{figure}

\begin{figure}[htbp]
\centering{
    \subfloat[]{\includegraphics[width=0.45\textwidth]{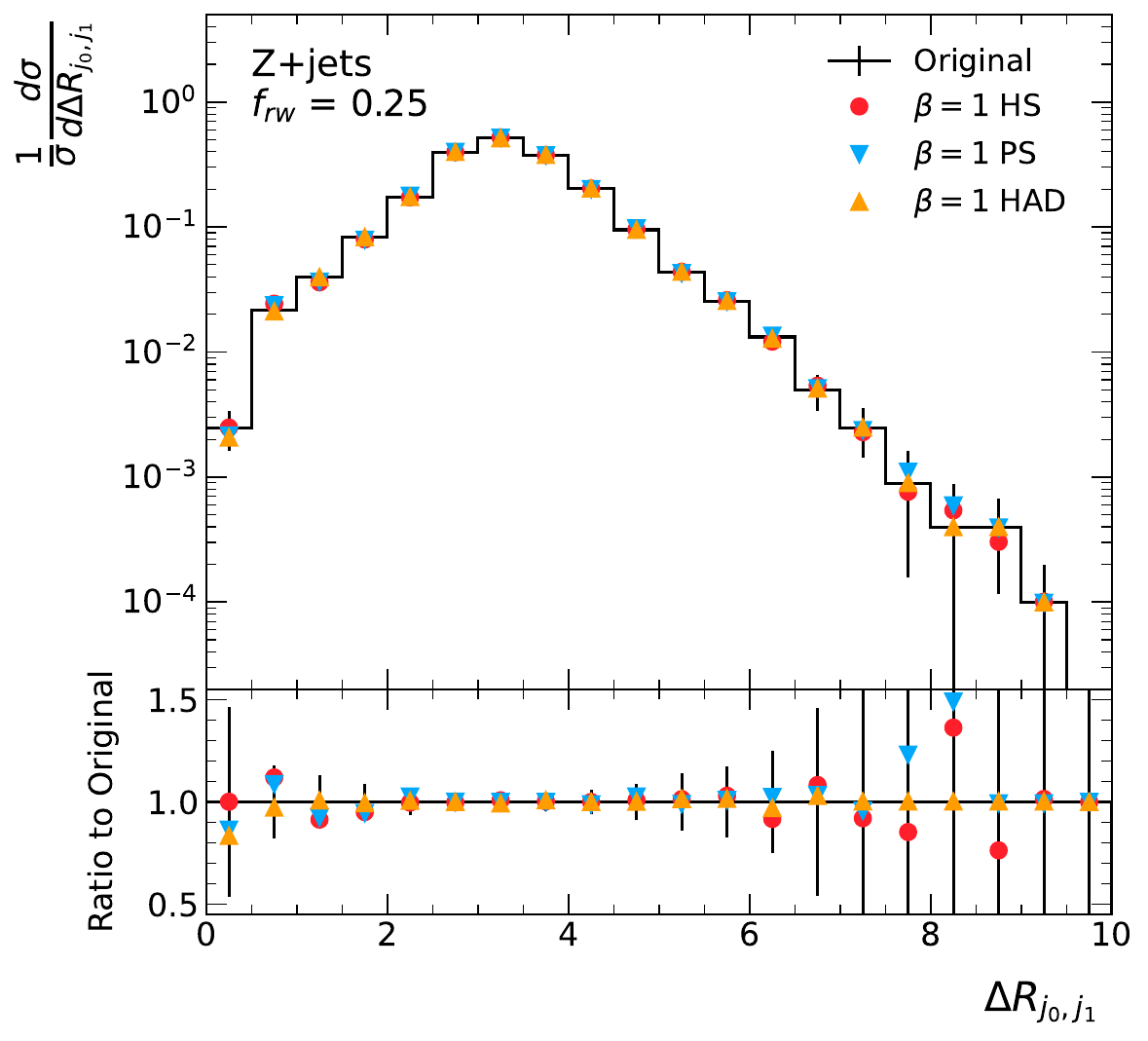}\label{fig:stages:drjj:25}}
   \subfloat[]{\includegraphics[width=0.45\textwidth]{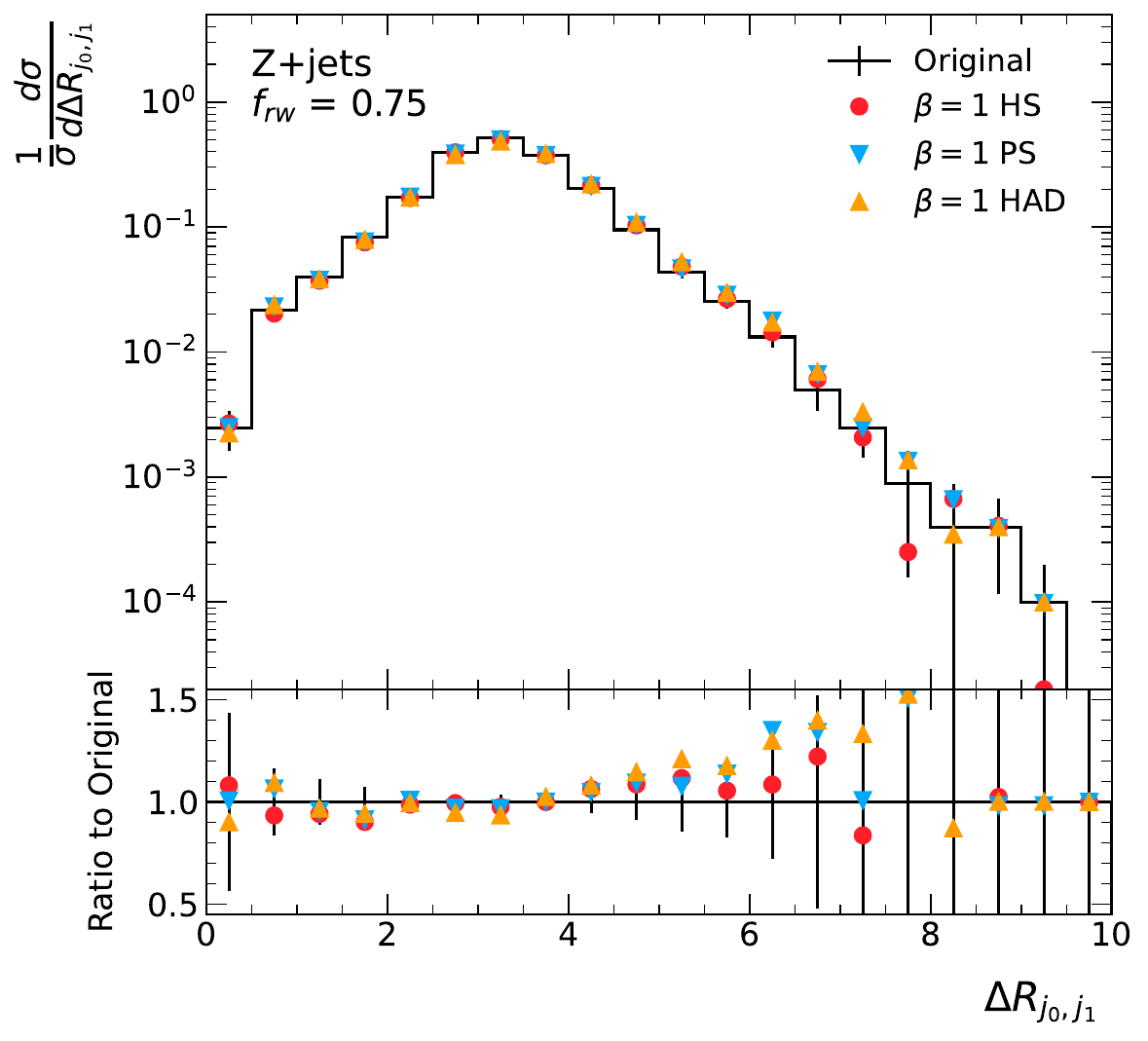}\label{fig:stages:drjj:75}}\\
   \subfloat[]{\includegraphics[width=0.45\textwidth]{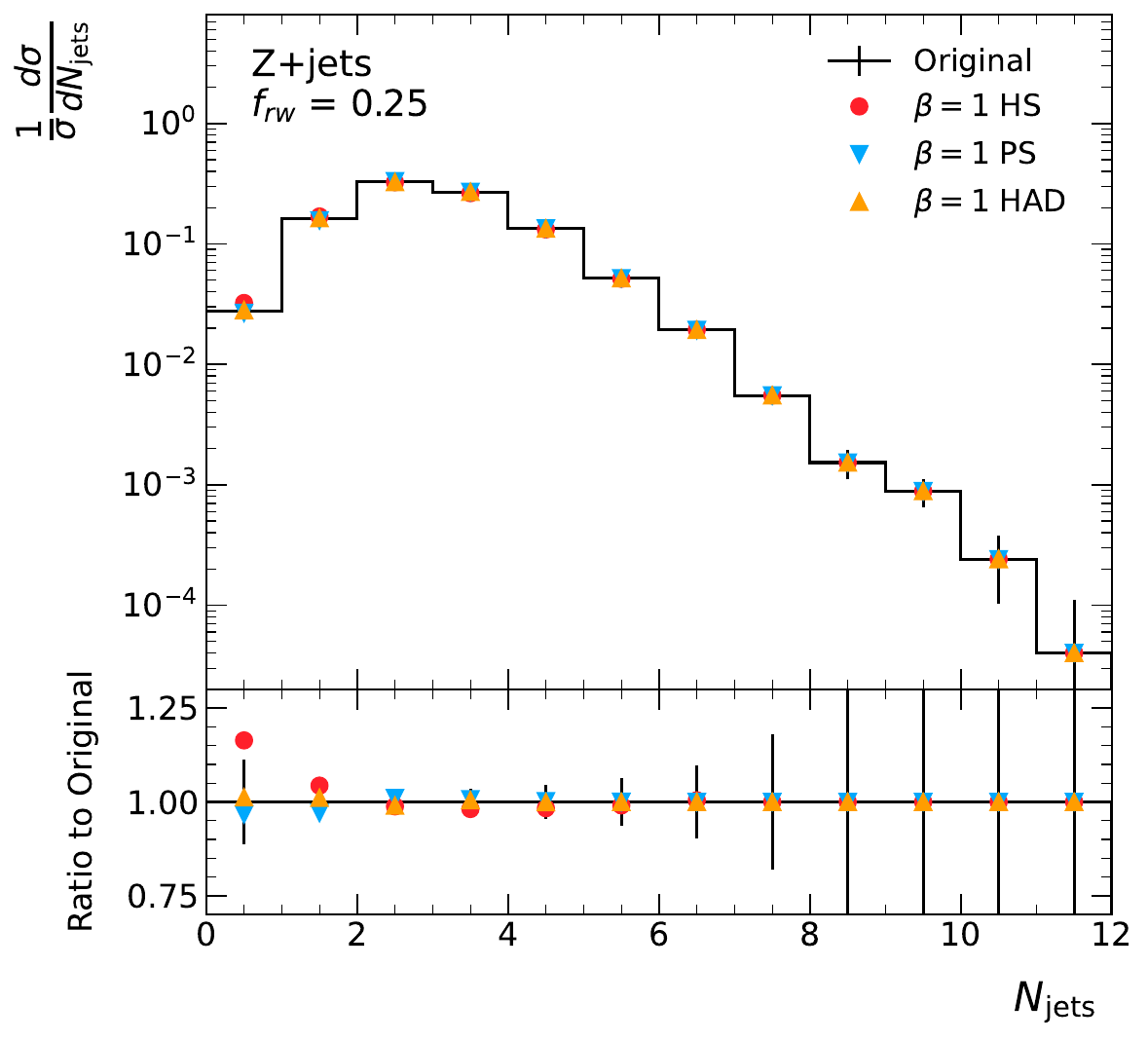}\label{fig:stages:njets:25}}
    \subfloat[]{\includegraphics[width=0.45\textwidth]{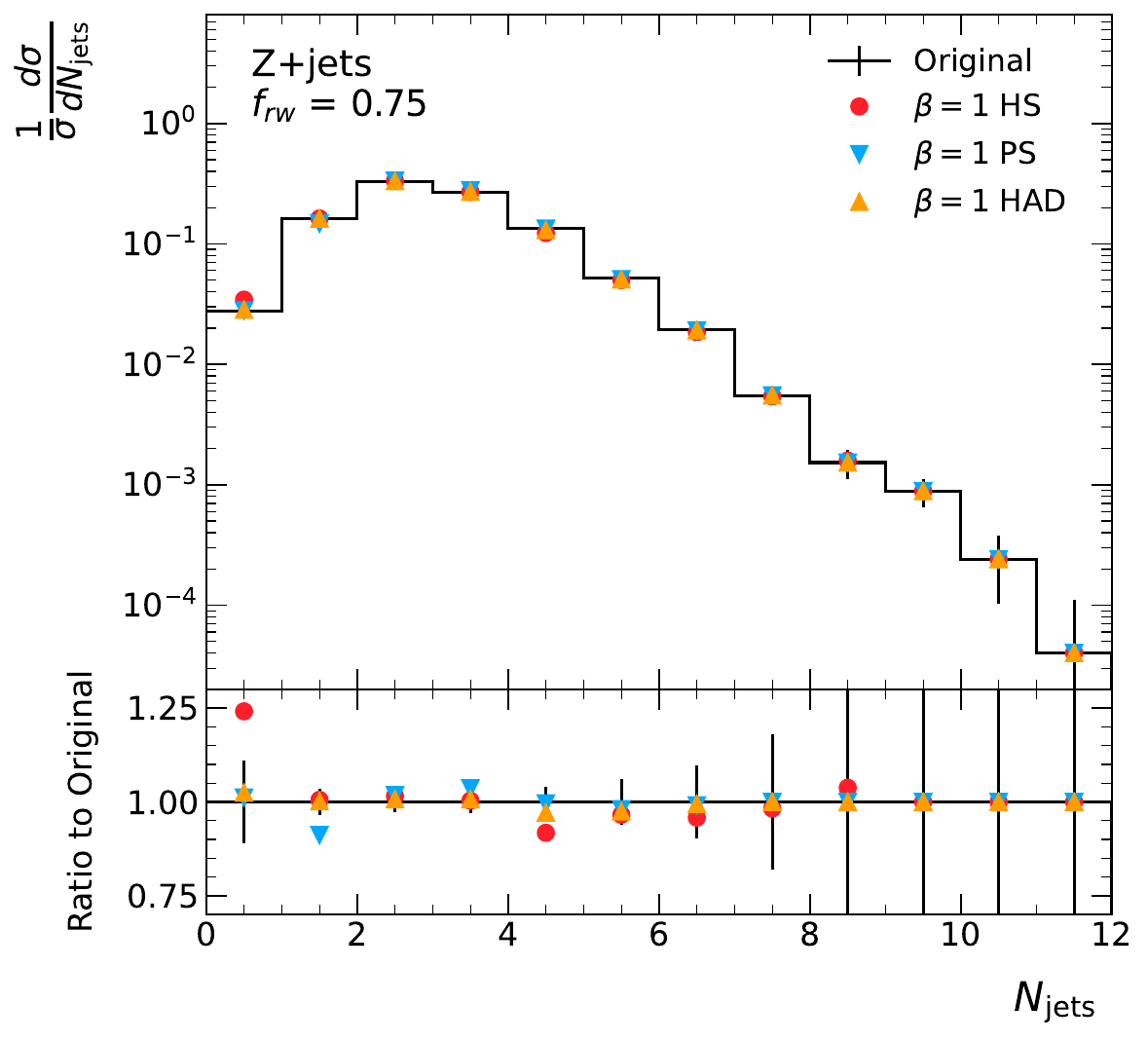}\label{fig:stages:njets:75}}
   \caption{Comparison of the normalized hadron-level unweighted \zjets  (\ref{fig:stages:drjj:25}, ~\ref{fig:stages:drjj:75}) \drjj, (\ref{fig:stages:njets:25} and ~\ref{fig:stages:njets:75}) \njets distributions to samples with \frw$=0.25$ (left) and \frw$=0.75$ (right) using HS, PS, and HAD event weights with $\beta=1$.}
 \label{fig:stages:beta1:2}
 }
\end{figure}

Figures~\ref{fig:stages:beta1:1}--\ref{fig:stages:beta1:2} show these three reweighting options for the observables outlined in Sec.~\ref{sec:fom:observables} in \zjets events, with a maximum cell radius chosen such that \frw is either 0.25 or 0.75.
The reweighting is performed on a sample with $10^5$ events, with the ratio to the original event sample shown as a comparison.
The distributions have been normalized, as the reweighting does not change the overall cross section.
Given the relatively small reweighted sample size, the bias produced by the cell reweighting procedure is exaggerated relative to more realistic scenarios, as the cell size decreases with the number of events.
For \frw =0.25, all three setups are compatible with the original samples for most bins in the distributions, with the exception of the jet multiplicity distribution in the HS reweighting, which disagrees with the fraction of events with no reconstructed jets by 15\%, slightly beyond the statistical uncertainty of the original sample. 

For the larger \frw =0.75, the differences between the reweighting strategies become more apparent.
All reweighting strategies agree with the \drjj and \ptrat distributions within statistical uncertainties.
The HS reweighting shows a significant bias in the \njets distribution, particularly in the lowest bin where it overestimates the cross section, and for intermediate values of $4<$\njets$<5$.
It also shows some tensions for intermediate \Ht values around 150 GeV, though these effects are smaller.
The HAD reweighting generally shows the most stable performance, with the exception of \drjj, where it systematically overestimates the cross section in the large-angle region.
Given the large statistical uncertainties in this region and the fact that this produces a more smoothly falling spectrum, this is not a cause of concern.
The PS reweighting shows less bias than the HS for the number of jets, but a larger bias than the HS for \Ht for the smallest \Ht bin.
On the whole, the hadronization-based reweighting shows the best performance for this set of observables.

\begin{figure}[htbp]
\centering{
\includegraphics[width=0.67\textwidth]{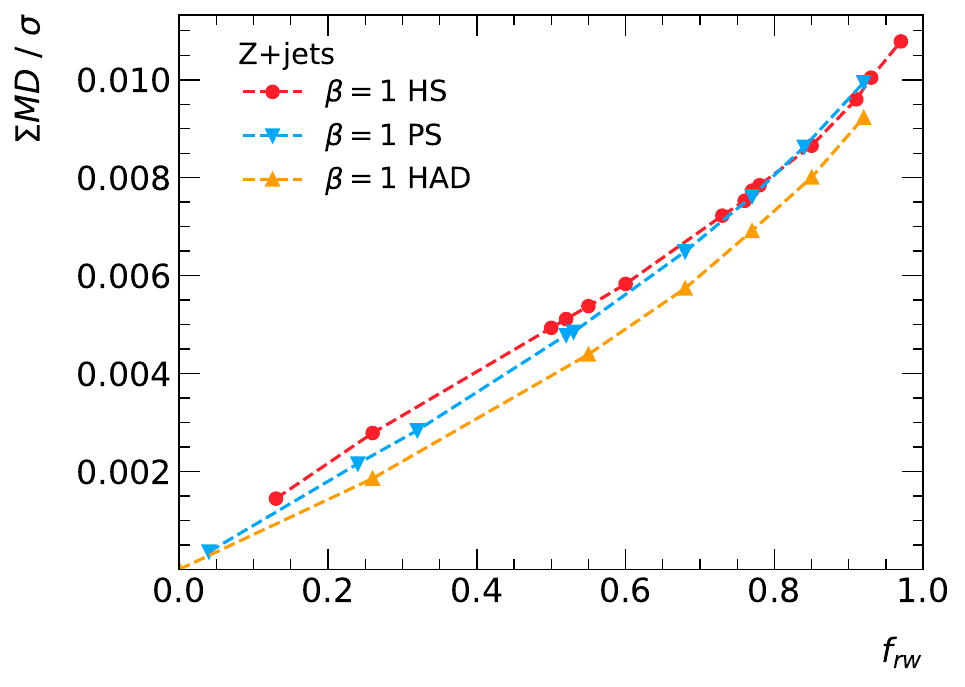}
   \caption{A comparison of the \XMD for the events where the reweighting is performed based on the hard process-, parton shower-, and hadronization-level information.}
    \label{fig:xmd:stages}
    }
\end{figure}

Figure~\ref{fig:xmd:stages} shows the \XMD for the HS, PS and HAD reweighting setups as a function of the fraction of negatively-weighted events that have been reweighted.
The HAD reweighting consistently has a lower \XMD with respect to the original weighted sample than the HS and PS setups.
For \frw below 0.7, the PS reweighting falls between the HS and HAD reweighting, while above 0.7, its \XMD is consistent with that of the HS reweighting.
These results are consistent with the behavior observed for individual kinematic distributions. 
As it provides the best performance, studies are performed on the fully hadronized event for the remainder of this paper.

\FloatBarrier
\subsection{Comparison of different \texorpdfstring{$\beta$}{beta} values}\label{sec:results:betas}

\begin{figure}[htbp]
\centering{
\includegraphics[width=0.67\textwidth]{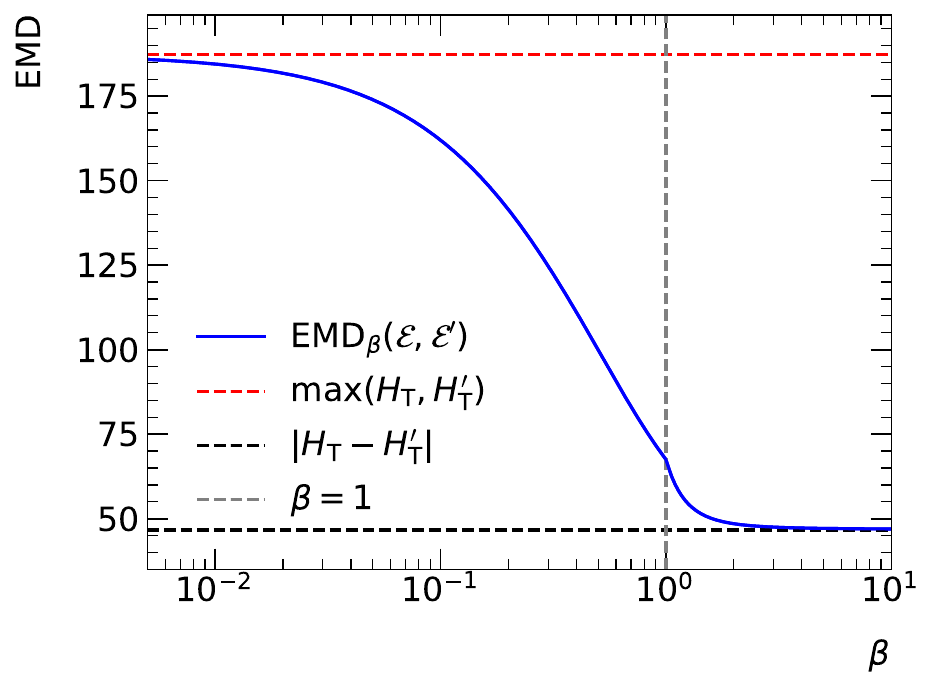}
   \caption{An illustration of the EMD between two events, $\mathcal{E}$ and $\mathcal{E^\prime}$, calculated with different choices of angular parameter $\beta$.
   The two events had respective \Ht values of 187~GeV and 141~GeV.}\label{fig:emd:betaScan}
   }
\end{figure}

The $\beta$ parameter controls the sensitivity of the EMD to transport with different angular scales; for larger values of $\beta$, energy can be transported through small angular displacements with less penalty.
The EMD between two events, $\mathcal{E}$ and $\mathcal{E^\prime}$, which have different (arbitrary) values of \Ht is shown in Figure~\ref{fig:emd:betaScan} for a continuous scan of $\beta$ values.
As the angular weight between these events decreases, their EMD is found to increase monotonically.
In the limit of $\beta\rightarrow \infty$, the distance between events is the difference in \Ht between the two events, while for $\beta = 0$, the distance between events is simply the maximum \Ht of the two events. 
The choice of $\beta=0$ is unique in that it creates a large degeneracy in the distance between events; a given negatively-weighted event will be equidistant from all events with a smaller value of \Ht. 
Due to this degeneracy, the seeding for $\beta=0$ is ordered by \Ht in order to avoid a strong dependence of the weight distribution on the order of seed selection.

\begin{figure}[htbp]
\centering{
   \subfloat[]{\includegraphics[width=0.45\textwidth]{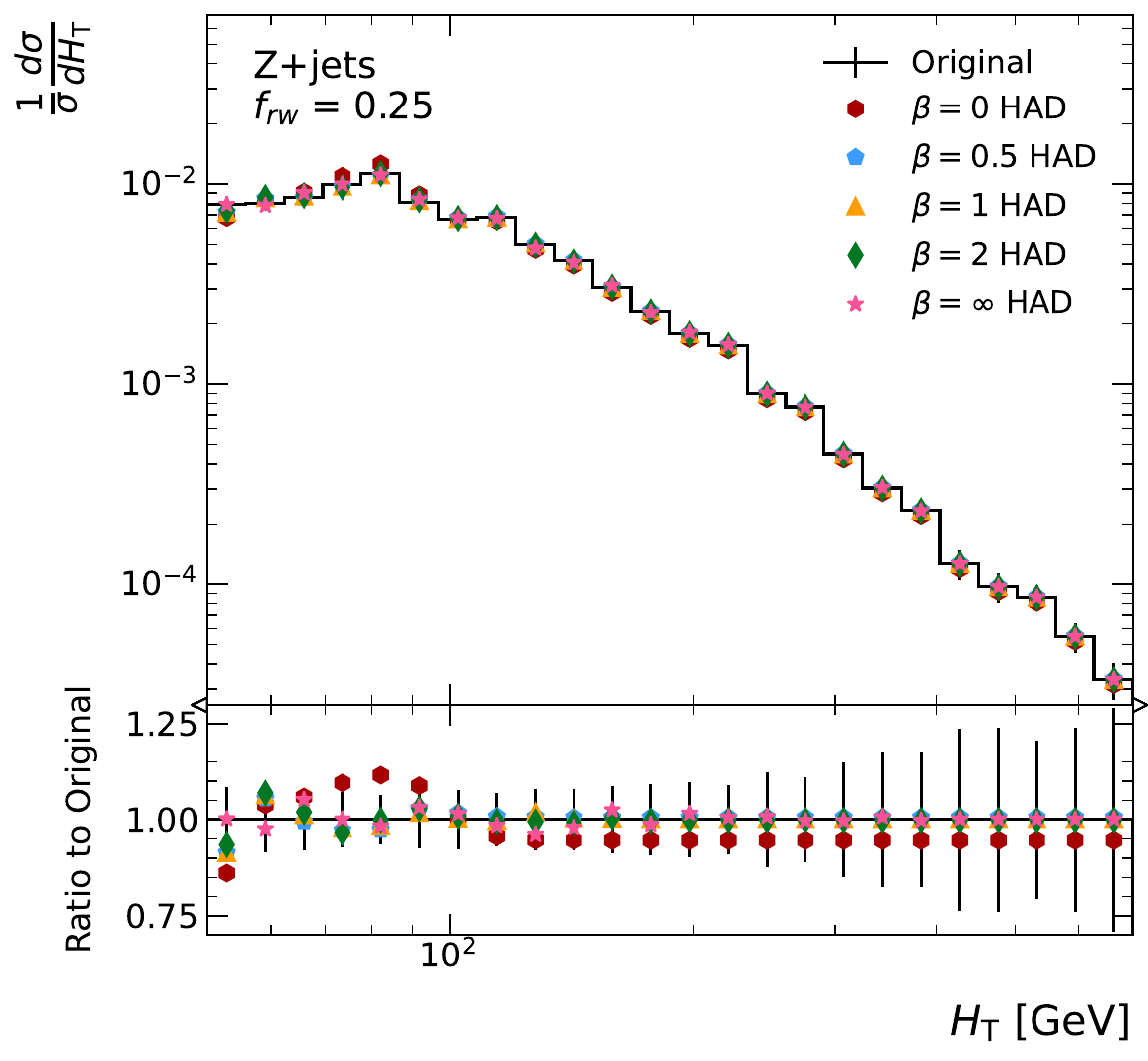} \label{fig:betas:ht:25}}
   \subfloat[]{\includegraphics[width=0.45\textwidth]{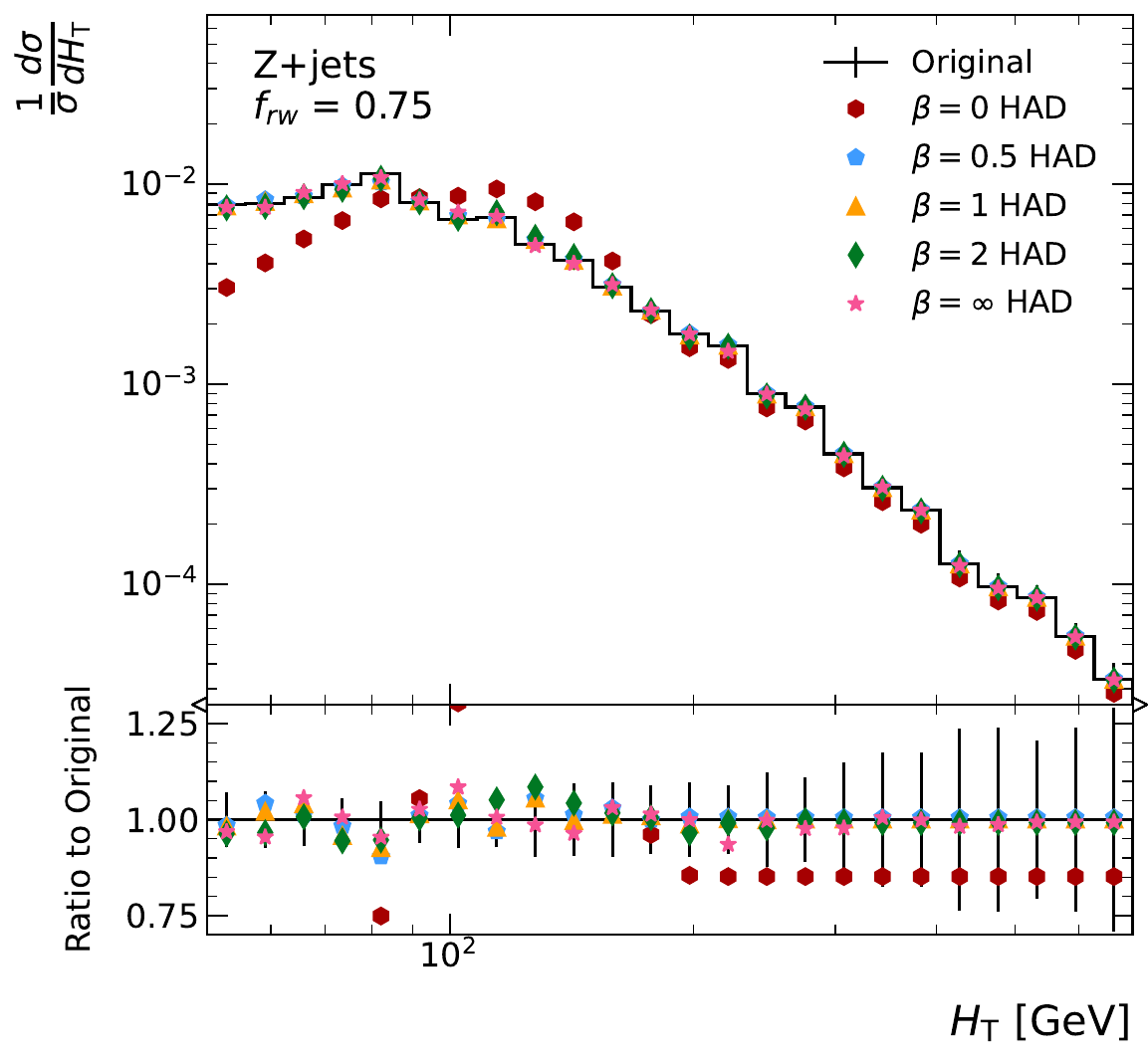} \label{fig:betas:ht:75}}\\
   \subfloat[]{\includegraphics[width=0.45\textwidth]{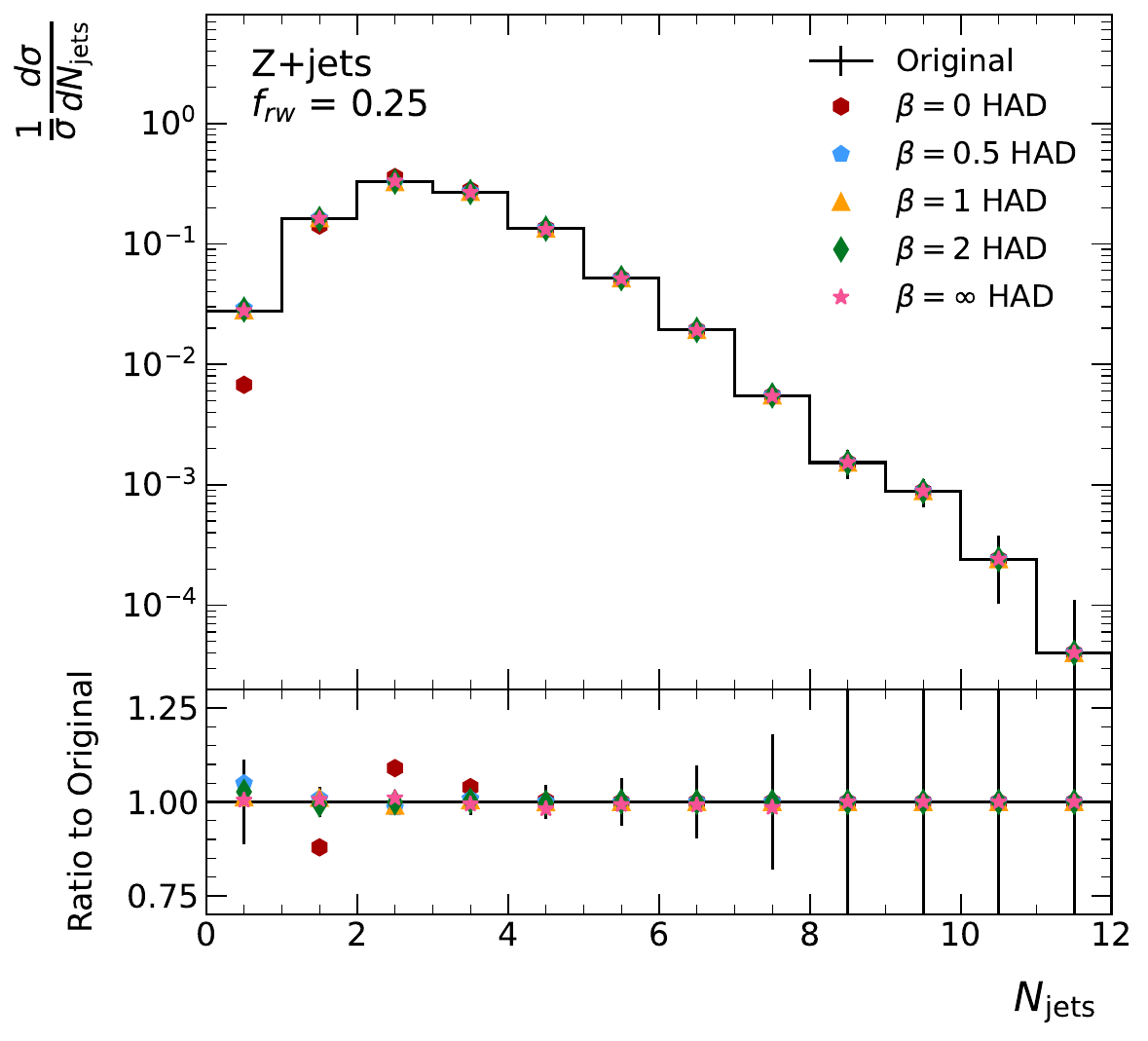} \label{fig:betas:njets:25}}
   \subfloat[]{\includegraphics[width=0.45\textwidth]{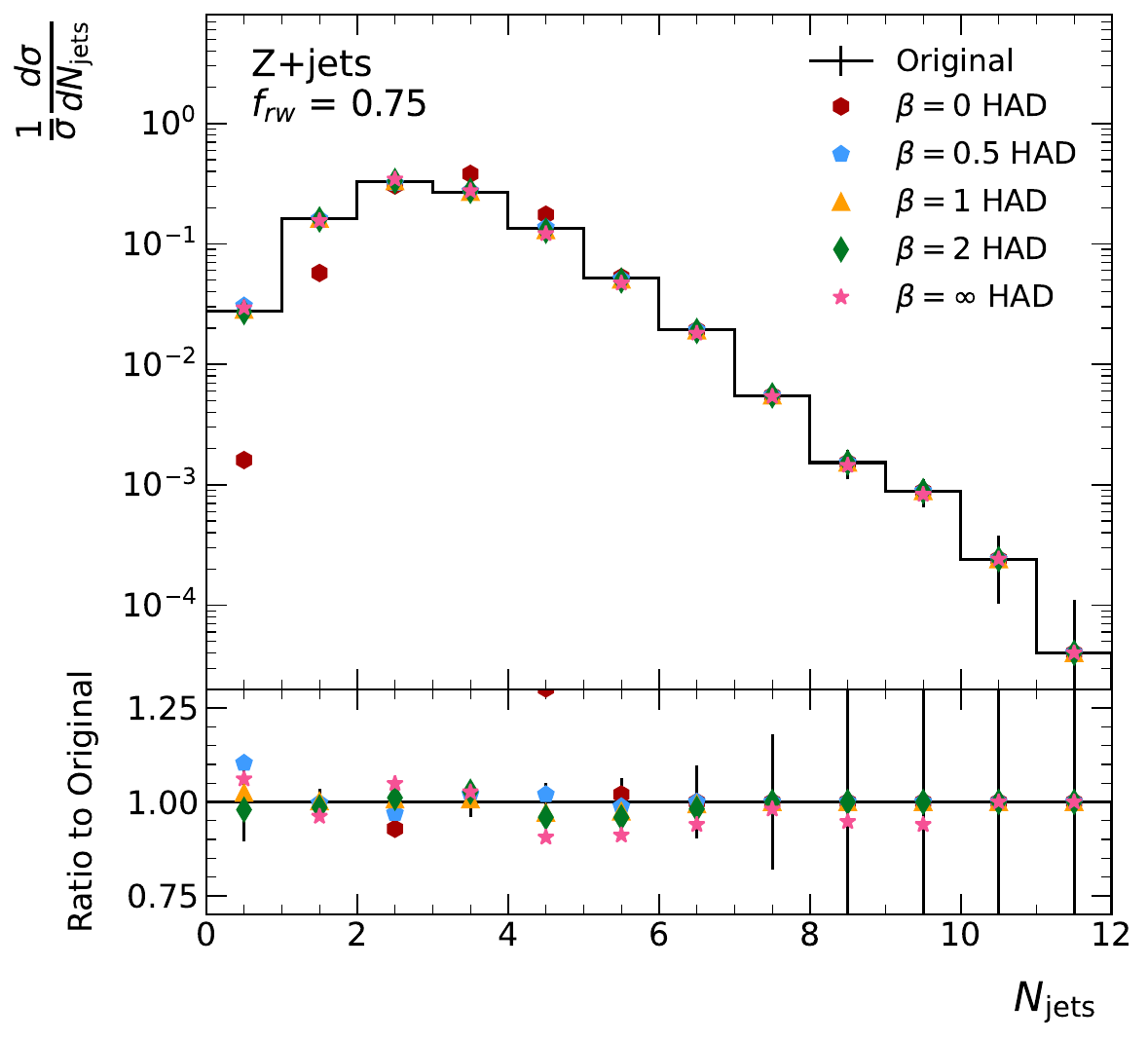} \label{fig:betas:njets:75}}

   \caption{Comparison of hadron-level unweighted \zjets  (\ref{fig:betas:ht:25}, ~\ref{fig:betas:ht:75}) \Ht and (\ref{fig:betas:njets:25}, \ref{fig:betas:njets:75}) \njets distributions to samples with \frw$=0.25$ (left) and \frw$=0.75$ (right) using $\beta= 0, 0.5, 1, 2$ and $\infty$.}
 \label{fig:betas:had:1}
 }
\end{figure}

\begin{figure}[htbp]
\centering{
   \subfloat[]{\includegraphics[width=0.45\textwidth]{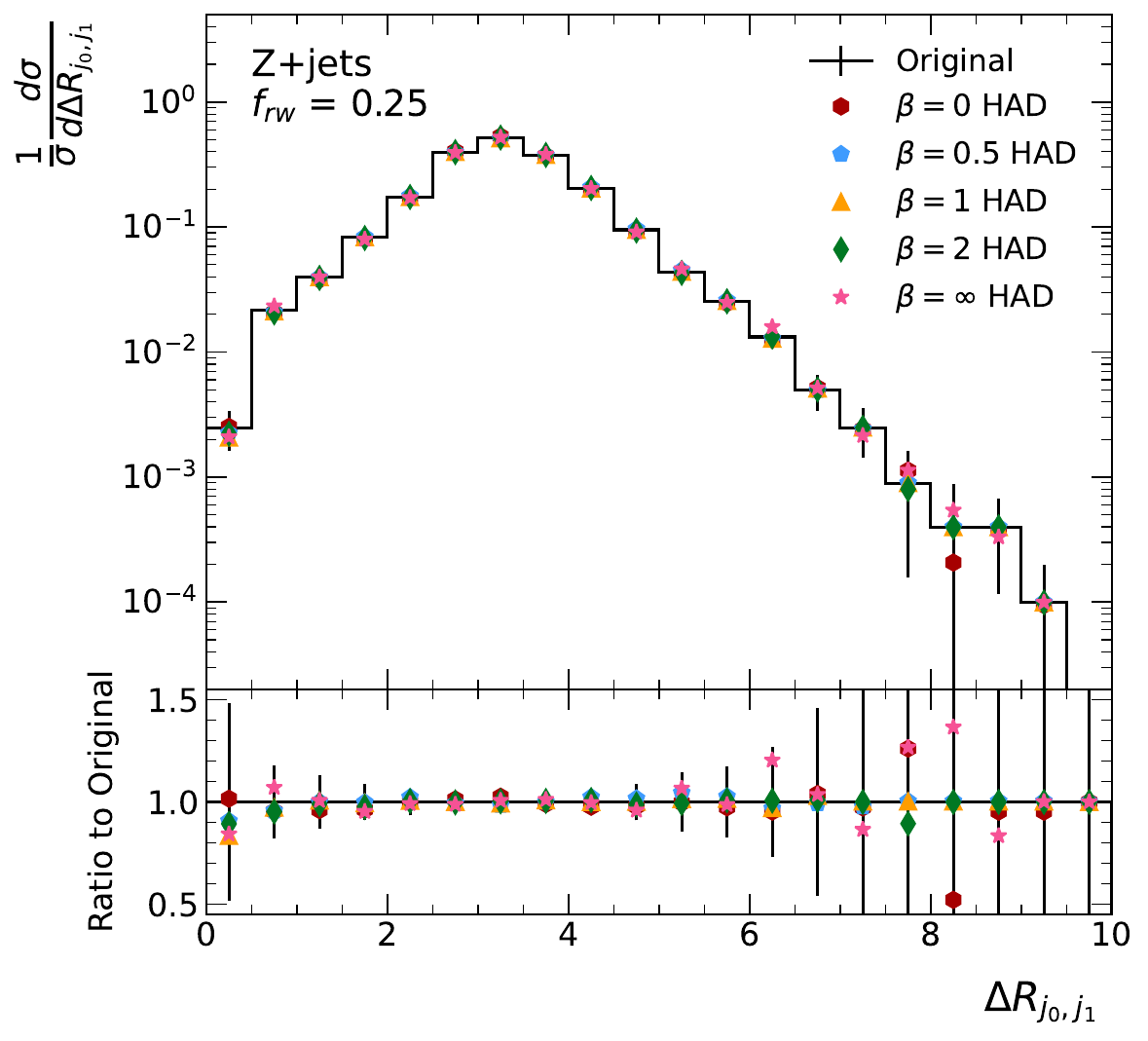} \label{fig:betas:drjj:25}}
   \subfloat[]{\includegraphics[width=0.45\textwidth]{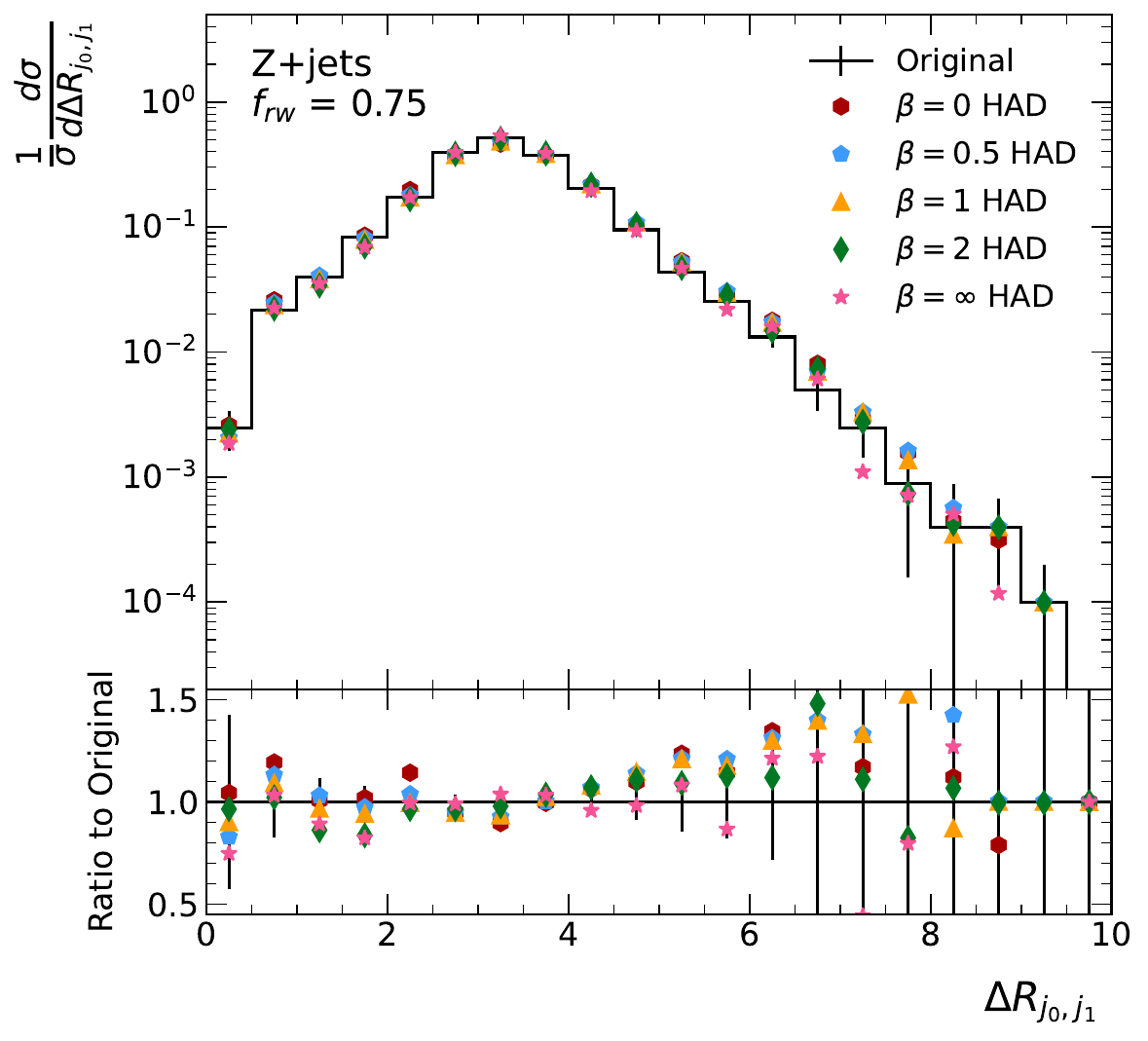} \label{fig:betas:drjj:75}}\\
   \subfloat[]{\includegraphics[width=0.45\textwidth]{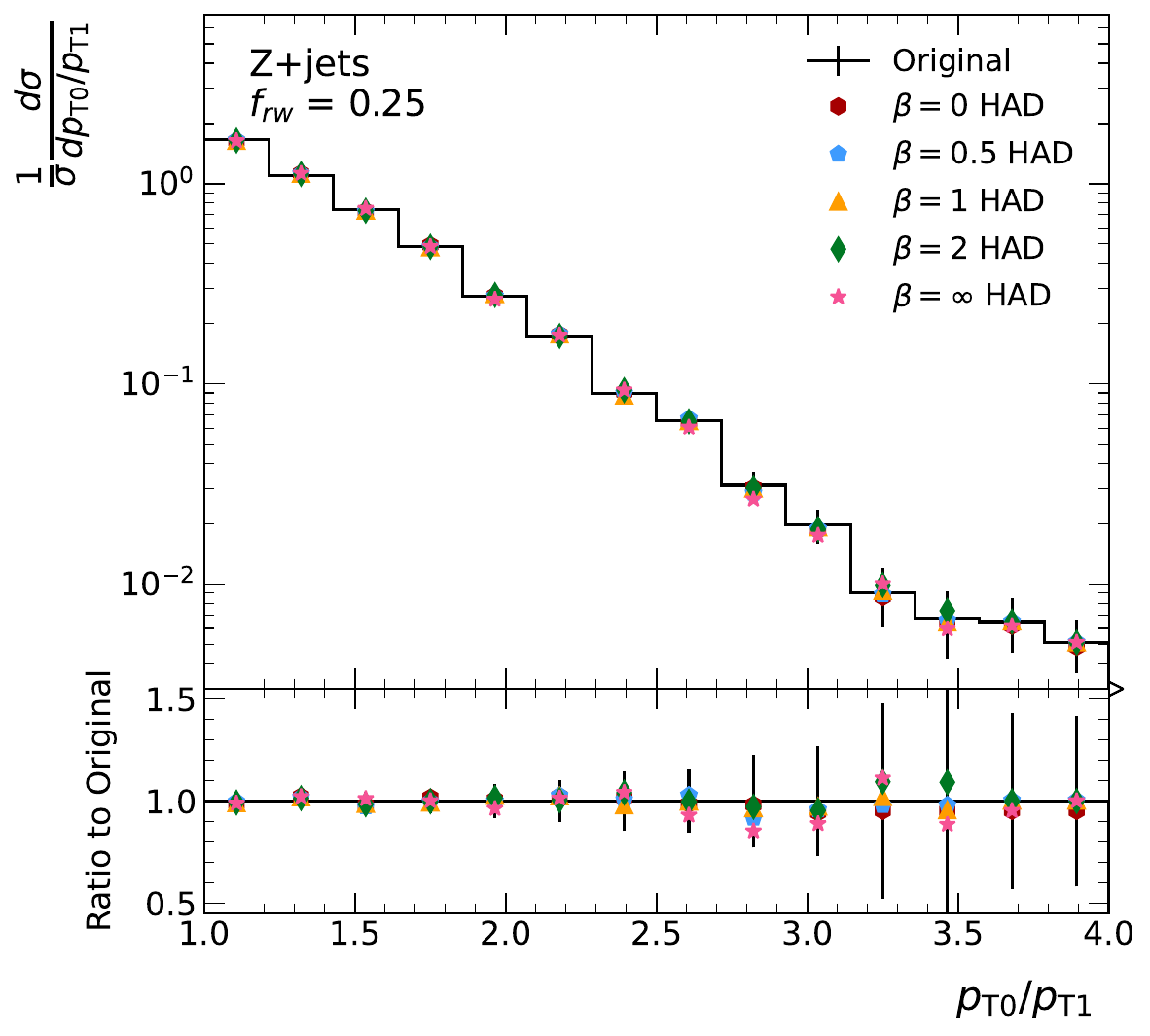} \label{fig:betas:ptrat:25}}
   \subfloat[]{\includegraphics[width=0.45\textwidth]{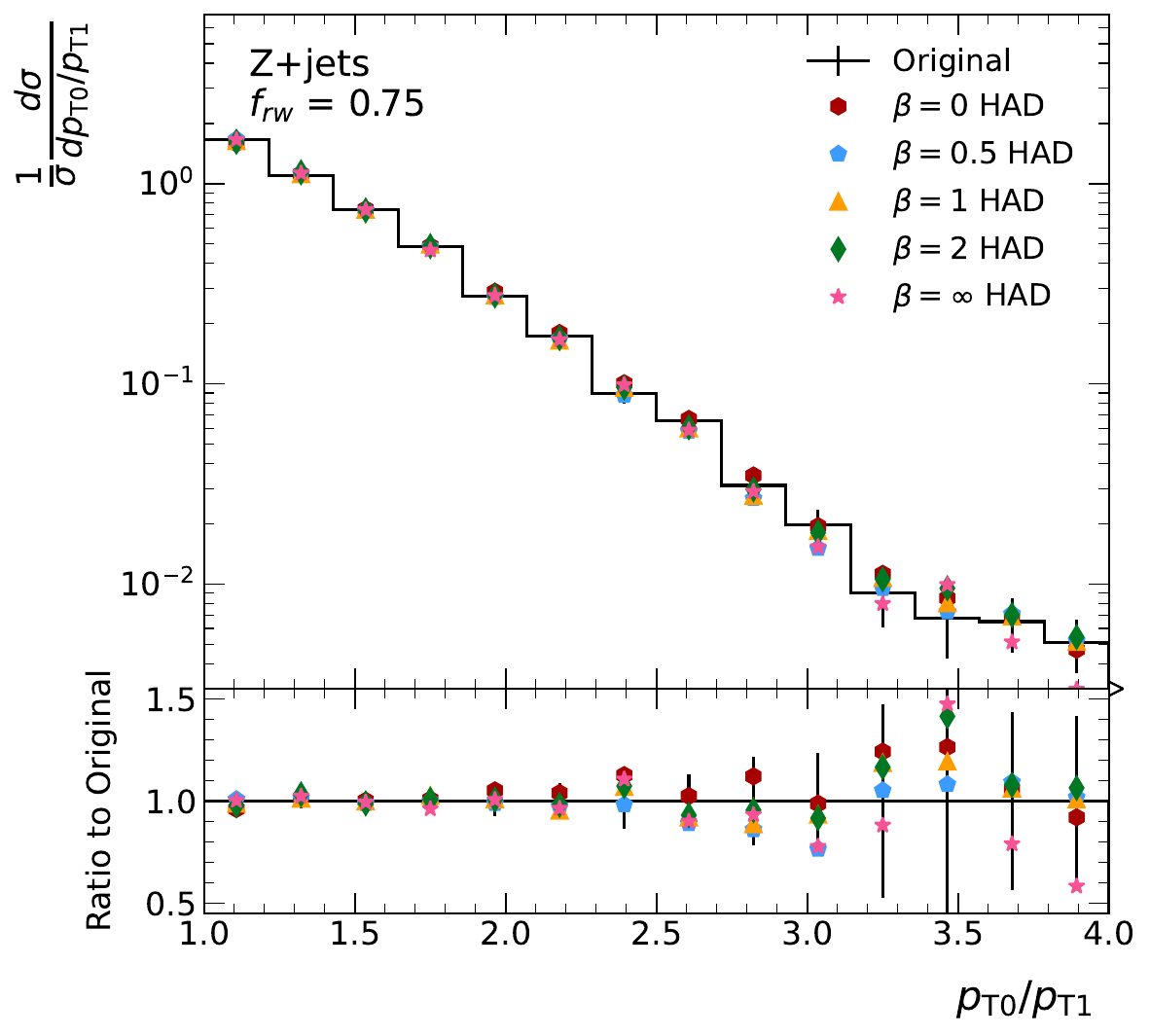} \label{fig:betas:ptrat:75}}
   \caption{Comparison of hadron-level unweighted \zjets   (\ref{fig:betas:drjj:25}, ~\ref{fig:betas:drjj:75}) \drjj and (\ref{fig:betas:ptrat:25}, ~\ref{fig:betas:ptrat:75}) \ptrat distributions to samples with \frw$=0.25$ (left) and \frw$=0.75$ (right) using $\beta= 0, 0.5, 1, 2$ and $\infty$.}
 \label{fig:betas:had:2}
 }
\end{figure}

The impact of varying the $\beta$ parameter during reweighting is seen in Figures~\ref{fig:betas:had:1}--\ref{fig:betas:had:2}, which show the kinematic dependence of the reweighting for $\beta=$ 0, 0.5, 1, 2, and $\infty$.
Even for the smaller \frw of 0.25, $\beta=0$ shows a measurable bias beyond the statistical uncertainty of the unweighted spectrum for \Ht $<100$ GeV and \njets $<5$.
For $\beta=0$, negatively-weighted events with small \Ht will produce the smallest cell radius, and will be reweighted first.
After reweighting, all of these event weights will decrease, creating a bias towards higher \Ht values.
This bias also appears in correlated observables such as \njets.
These biases worsen for \frw $= 0.75$, with deviations in the \Ht spectrum of nearly 50\%, well beyond the statistical precision of the original sample.
The $\beta \rightarrow \infty$ reweighting also has anomalous behavior, due to the fact that the reweighting is purely based on the \Ht difference between events.
For \frw$=0.25$, these deviations are within the statistical precision of the unweighted sample.
For \frw$=0.75$, the $\beta \rightarrow \infty$ reweighting has poor performance for the \njets distribution, with large deviations for $4 < $ \njets $< 6$.
For the other $\beta$ values, both \frw values of 0.25 and 0.75 have consistent performance with the original sample.

\begin{figure}[htbp]
    \centering
    \includegraphics[width=0.67\linewidth]{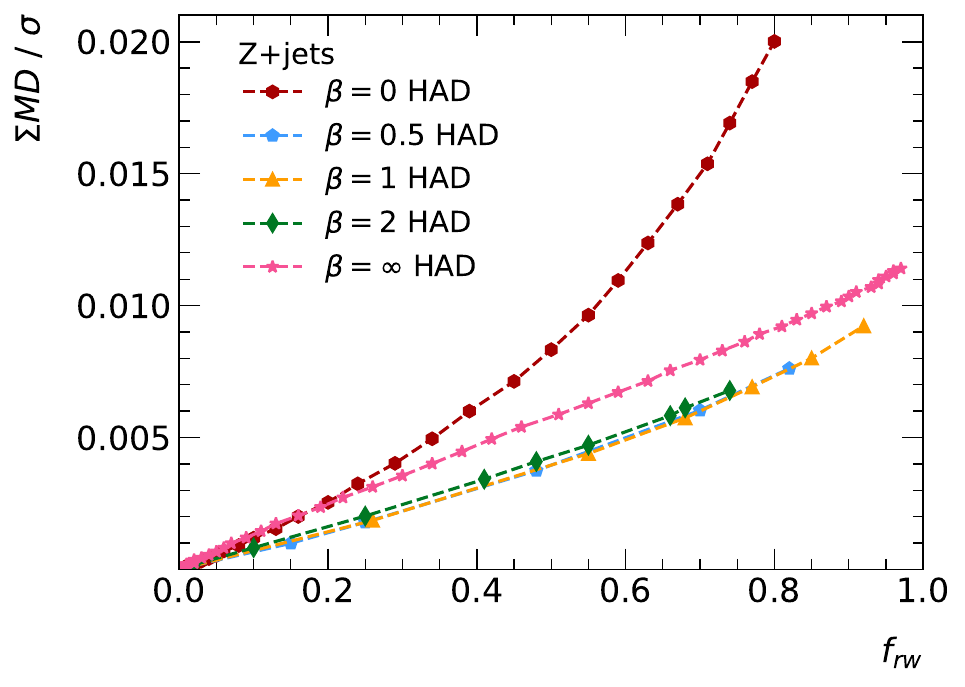}
    \caption{A comparison of the \XMD for $\beta=$ 0, 0.5, 1, 2, and $\infty$ computed from the hadronization-level information.}
    \label{fig:betas:xmd}
\end{figure}

\FloatBarrier
Figure~\ref{fig:betas:xmd} shows a comparison of the \XMD of all studied $\beta$ values.
The reweightings performed with $\beta=0$ and $\beta\rightarrow\infty$ have the largest \XMD compared to the original sample; their degraded performance is consistent with the 1D kinematic plots in Figures~\ref{fig:betas:had:1}--\ref{fig:betas:had:2} and occurs due to these values of $\beta$ failing to account for the spatial distribution of energy within events.
The $\beta=0.5$, $1$ and $2$ reweightings have comparable performance: $\beta=0.5$ and $\beta=1$ perform consistently for all values of \frw, while $\beta=2$ performs marginally worse.
Since the calculation for $\beta=0.5$ includes a square root, it is less computationally efficient (as discussed in Ref.~\cite{Cesarotti:2020xtf}).
The $\beta=1$ reweighting is therefore preferable and is the value selected for the remainder of these studies.

%%%%%%%%%%%%%%%%%%%%%%%%%%%%%%%%%%%%%%%%%%%%%%%%%%%%%%%%%%%%%%%%%%%%%%%%%%%%%%%

\FloatBarrier
\subsection{Comparison among metric choices}\label{sec:results:SEMD}

To study the impact of the metric itself on the reweighting, the optimal EMD configuration ($\beta=1$ applied to events after hadronization) is compared to the Spectral EMD (Sec.~\ref{sec:semd}) and to the Euclidean object-based reweighting developed by Andersen \emph{et al.} in Refs.~\cite{Andersen:2021mvw} and~\cite{Andersen:2024mqh}, including the separate treatment of leptonic and hadronic objects in the metric introduced in the latter reference.
An earlier version of the metric, where leptonic and hadronic objects are treated democratically, was also studied and was found to give similar results to those presented here.
Following the EMD optimization, the sEMD reweighting was studied only on the fully hadronized event.
These options represent three fundamentally different ways of constructing the event-wise distance metric, providing more insight into the intersection between the metric space and the stability of cell reweighting.

\begin{figure}[htpb]
\centering
   \subfloat[]{\includegraphics[width=0.45\textwidth]{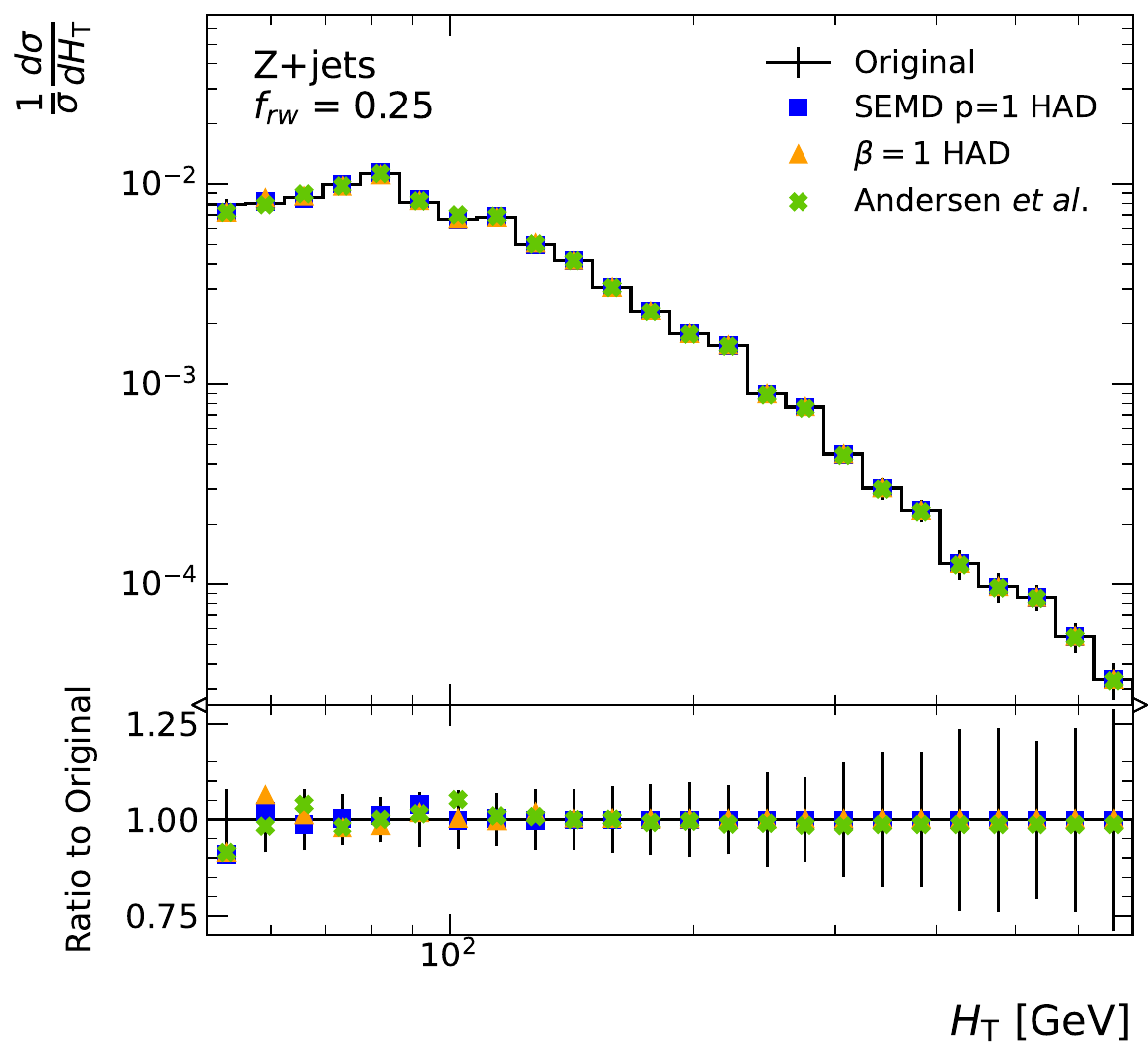}\label{fig:specter:had:ht:25}} 
   \subfloat[]{\includegraphics[width=0.45\textwidth]{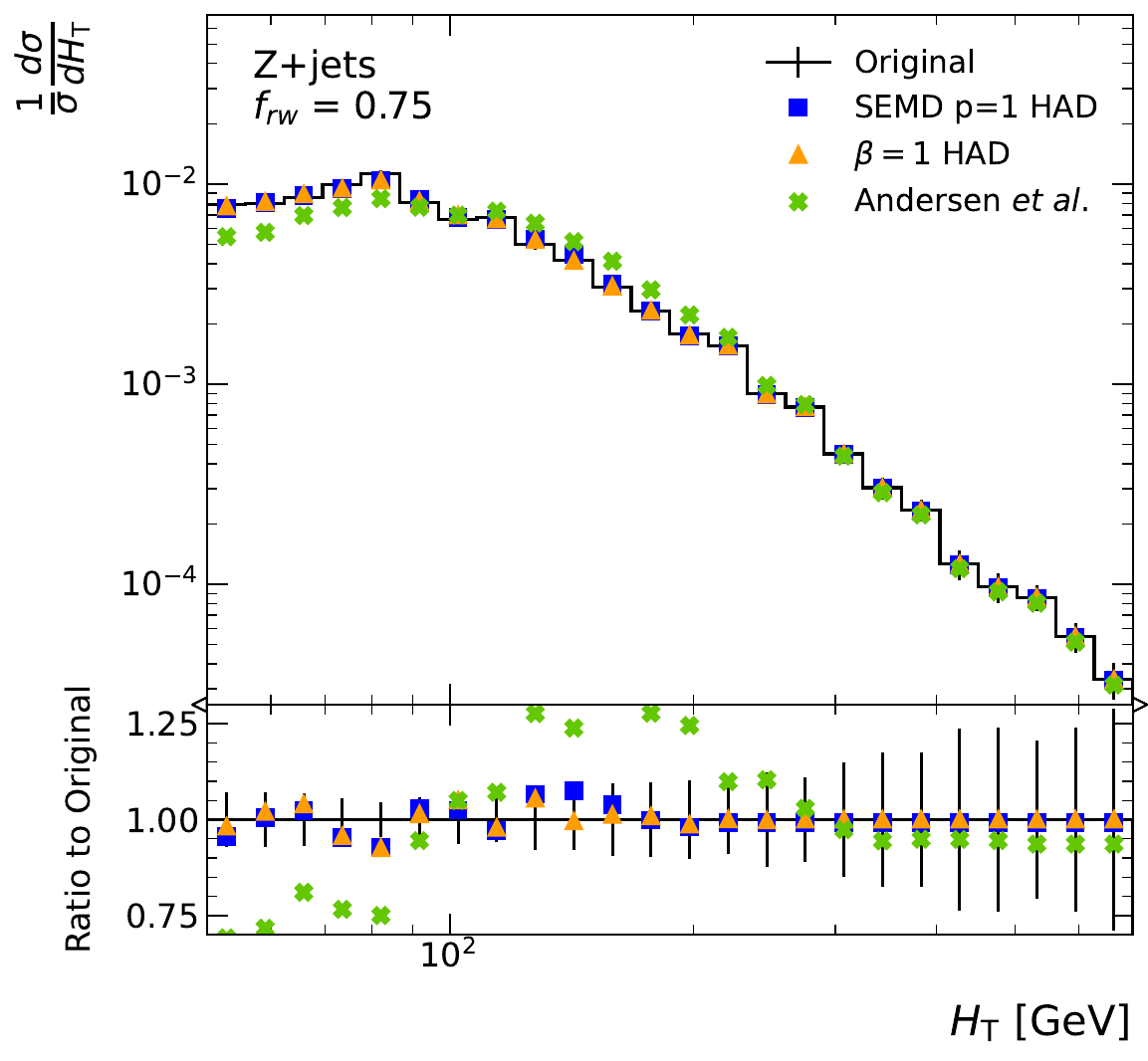}\label{fig:specter:had:ht:75}} \\
   \subfloat[]{\includegraphics[width=0.45\textwidth]{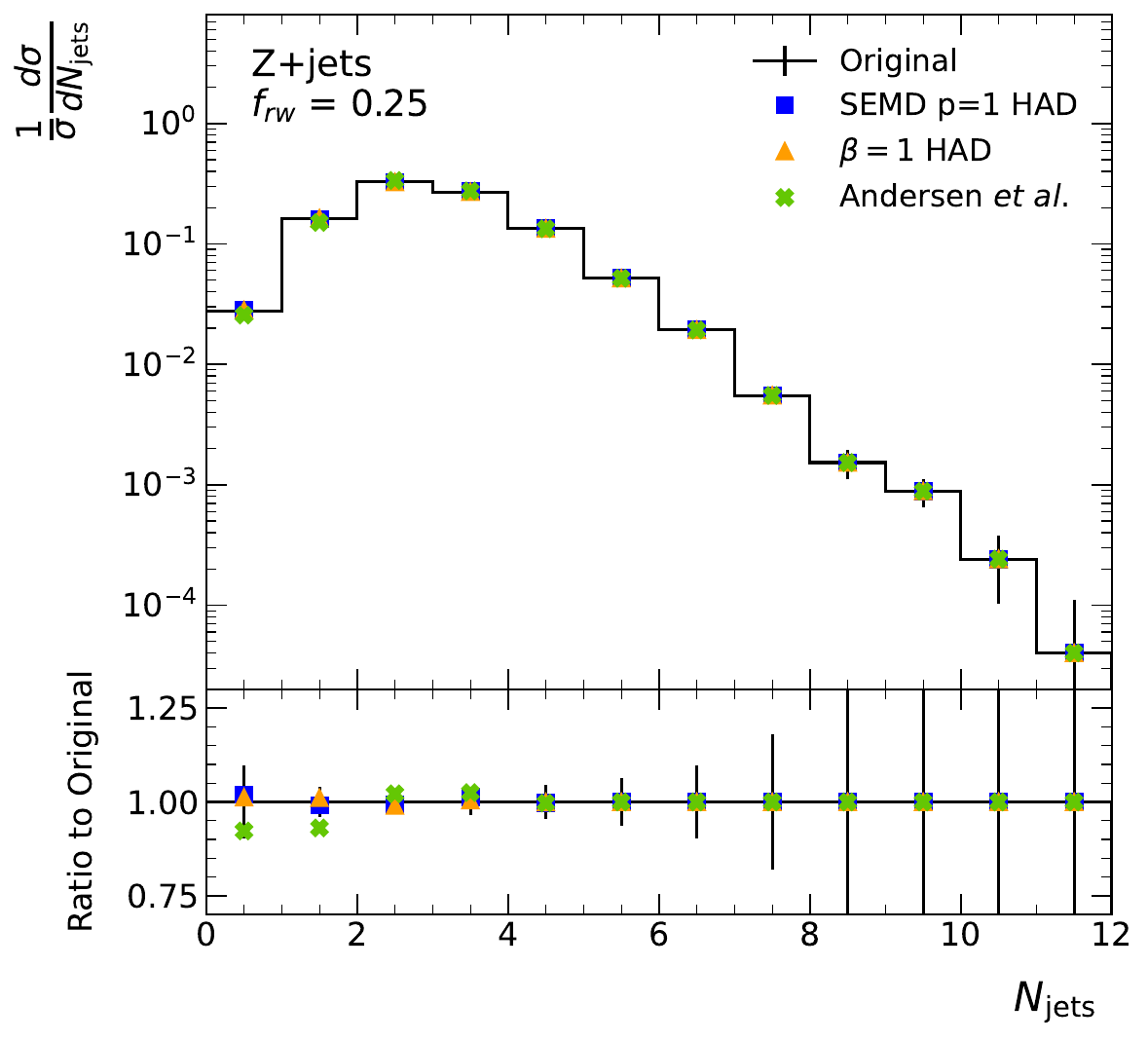}\label{fig:specter:had:njets:25}}   
   \subfloat[]{\includegraphics[width=0.45\textwidth]{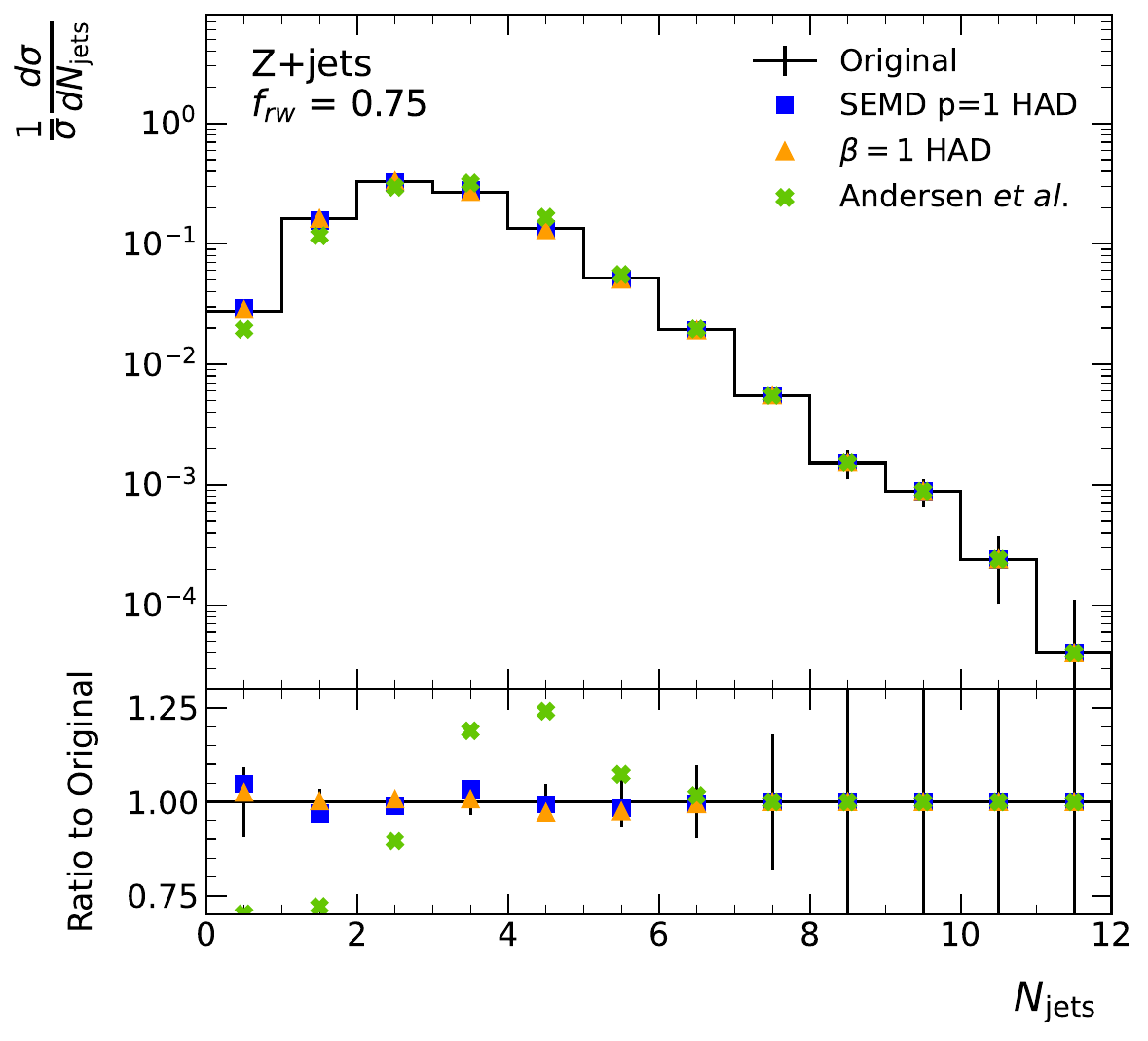} \label{fig:specter:had:njets:75} }
   \caption{Comparison of hadron-level unweighted \zjets  (\ref{fig:specter:had:ht:25}, ~\ref{fig:specter:had:ht:75}) \Ht and (\ref{fig:specter:had:njets:25}, \ref{fig:specter:had:njets:75}) \njets distributions to samples with \frw$=0.25$ (left) and \frw$=0.75$ (right) using EMDs and sEMDs at hadronization level.}
 \label{fig:specter:had:1}
\end{figure}

\begin{figure}[htpb]
\centering
   \subfloat[]{\includegraphics[width=0.45\textwidth]{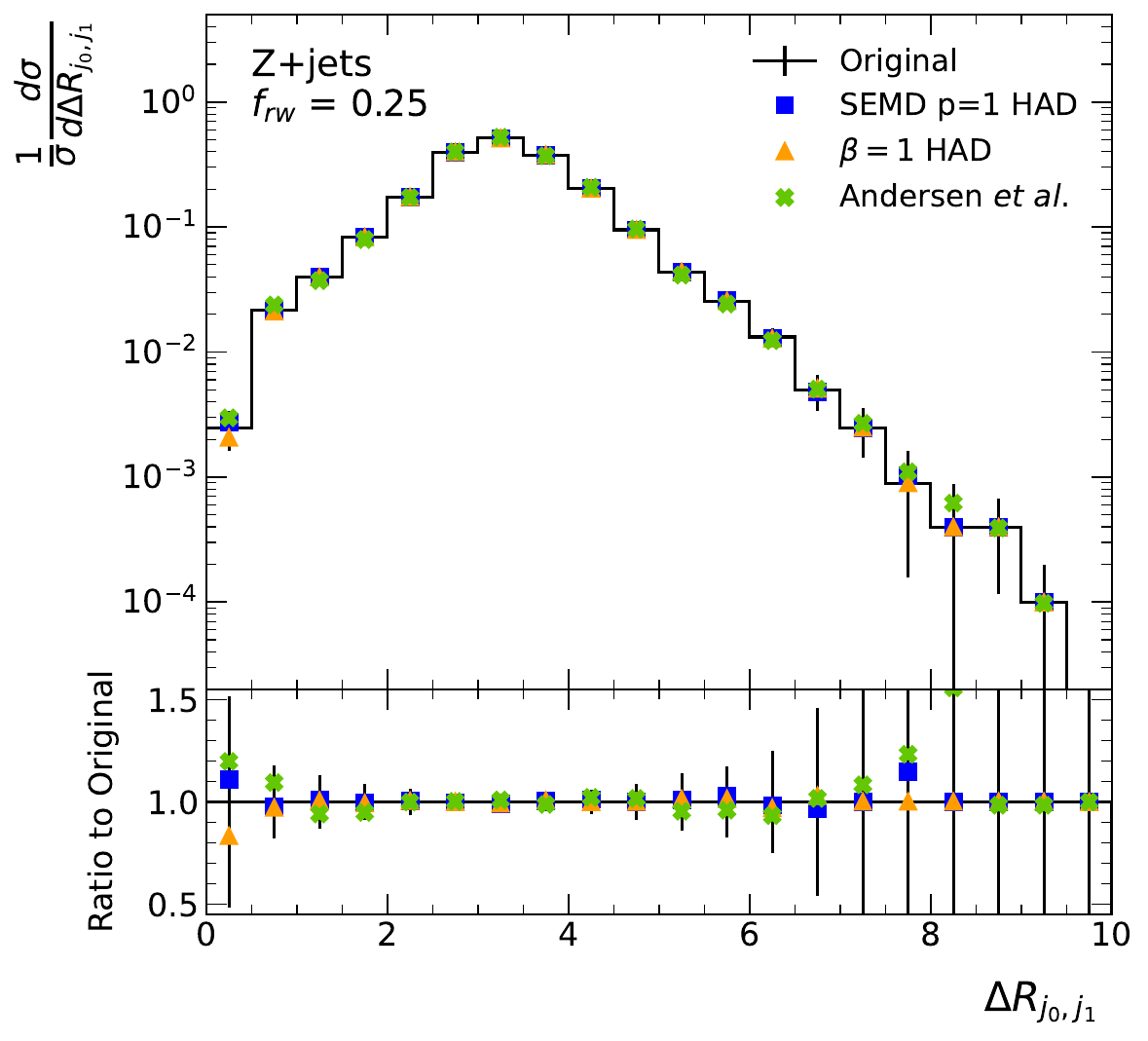}\label{fig:specter:had:drjj:25}} 
   \subfloat[]{\includegraphics[width=0.45\textwidth]{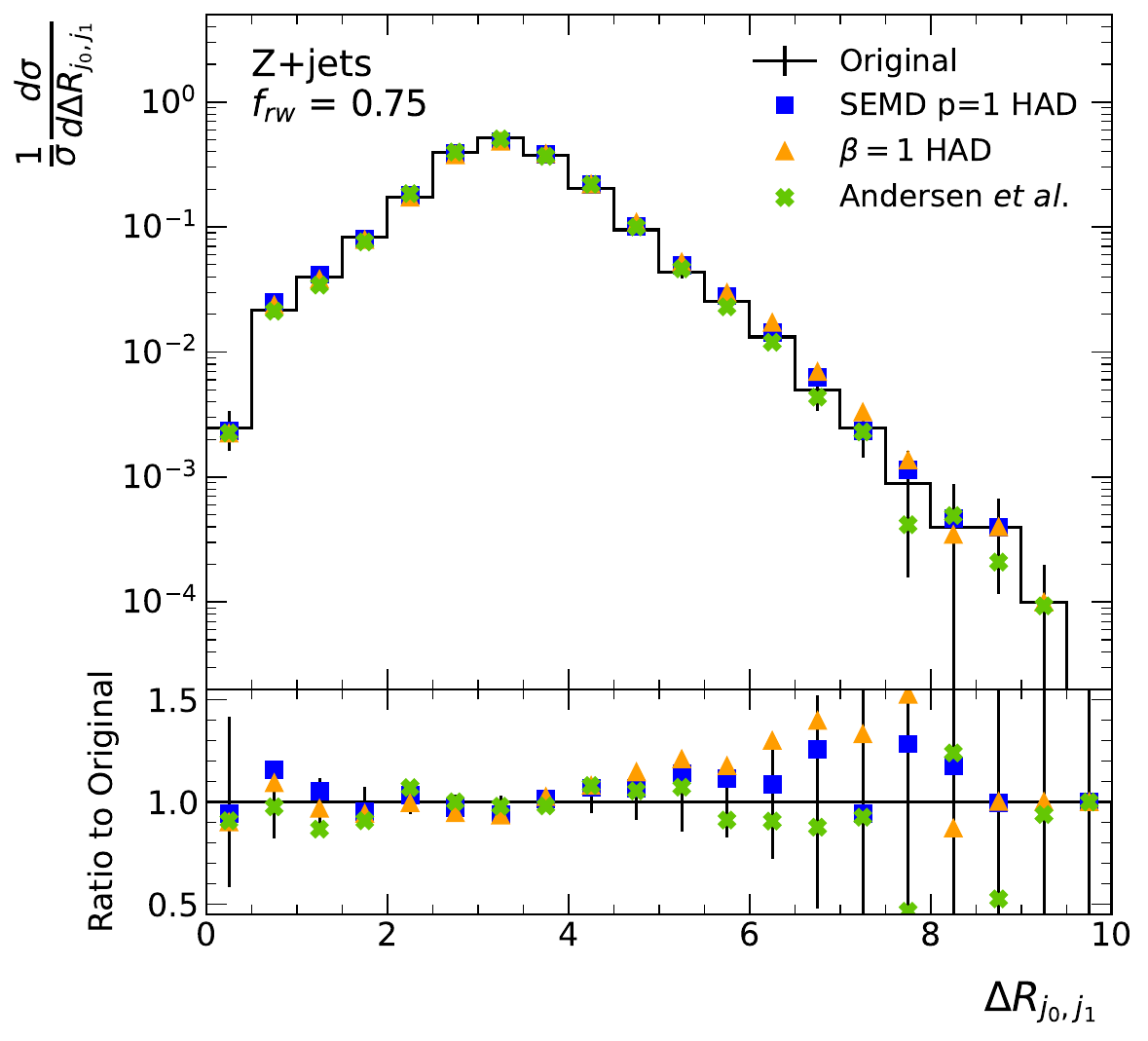}\label{fig:specter:had:drjj:75}}\\
   \subfloat[]{\includegraphics[width=0.45\textwidth]{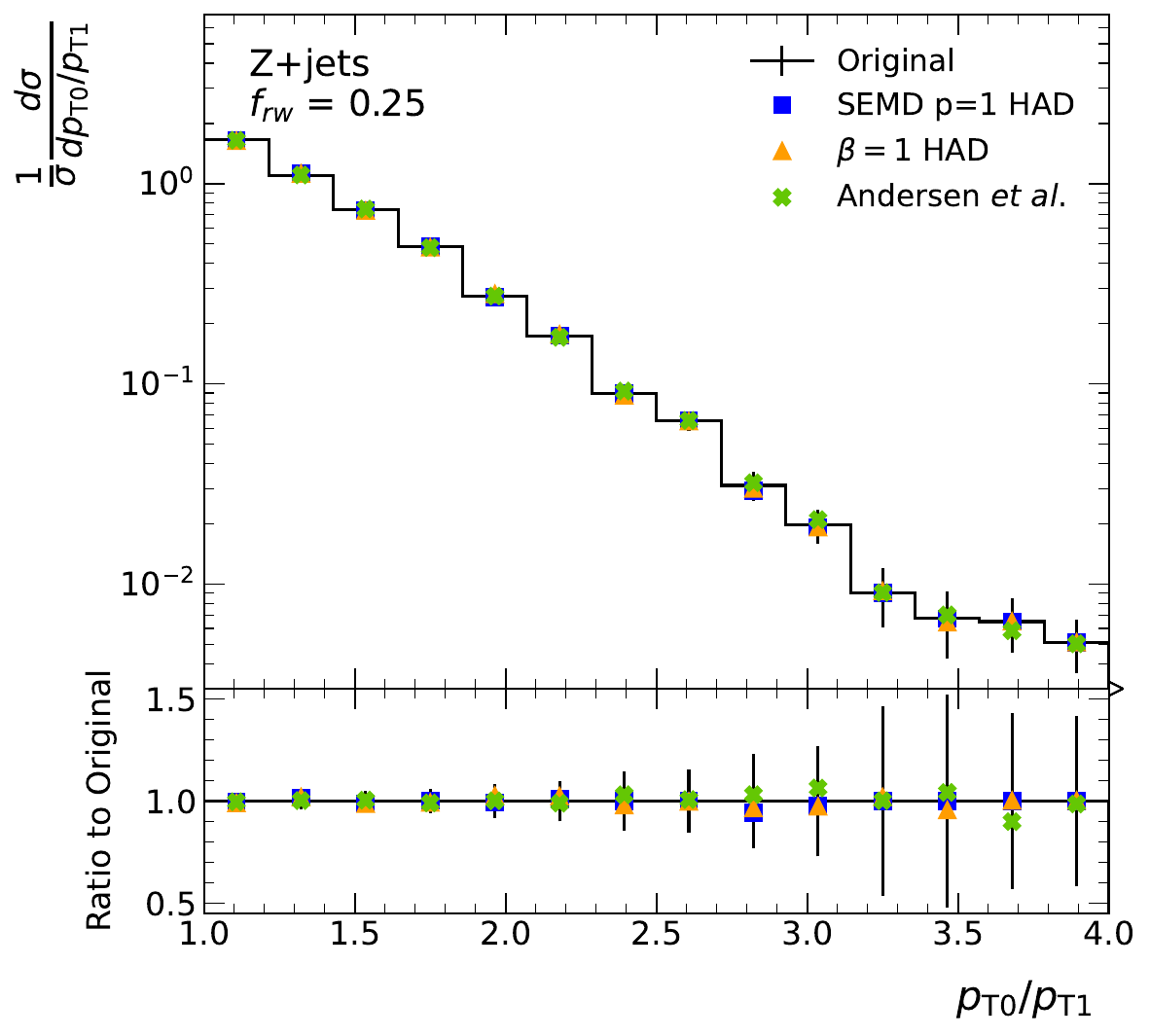}\label{fig:specter:had:ptrat:25}} 
   \subfloat[]{\includegraphics[width=0.45\textwidth]{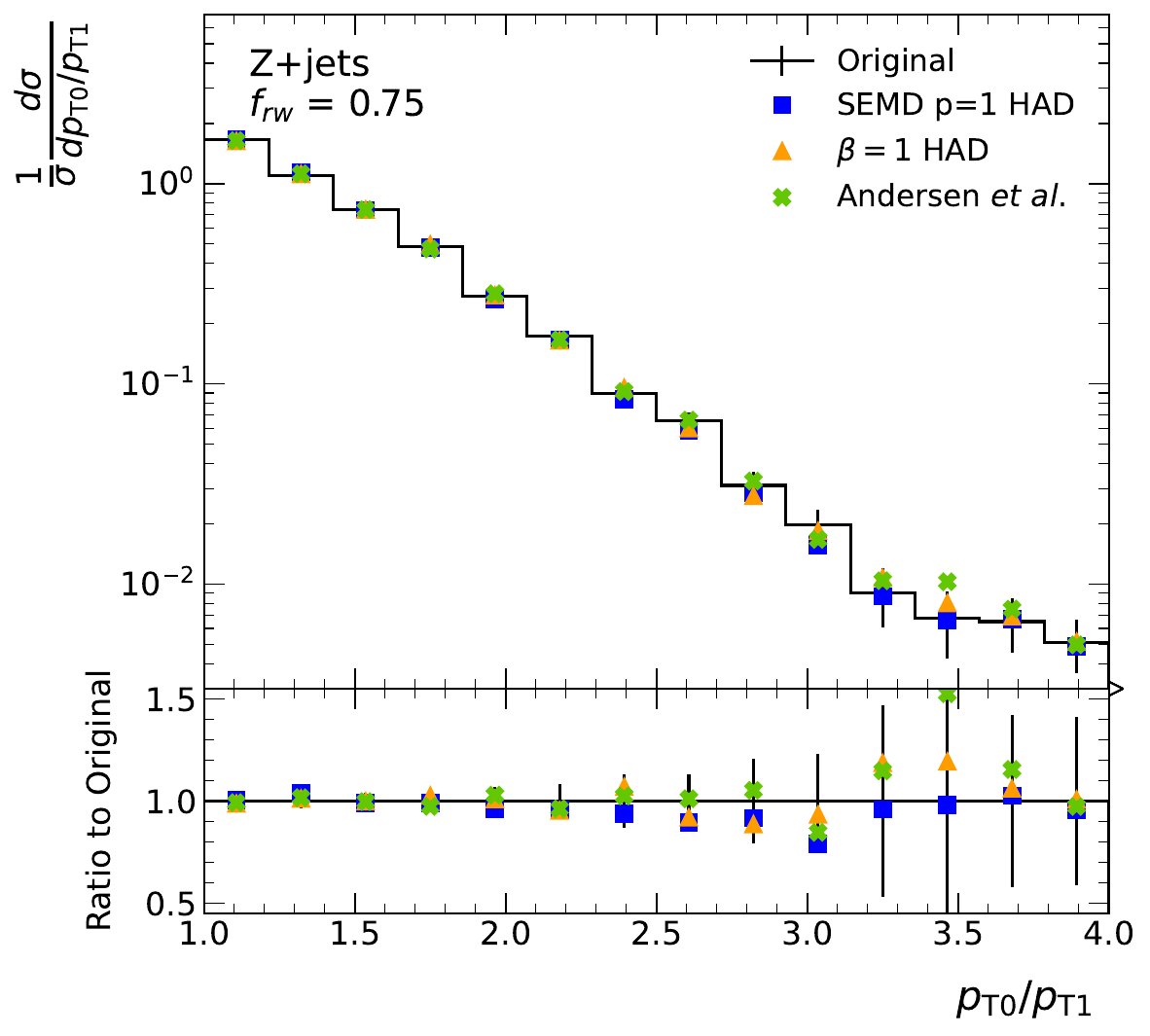}\label{fig:specter:had:ptrat:75}}
   \caption{Comparison of hadron-level unweighted \zjets  (\ref{fig:specter:had:drjj:25}, ~\ref{fig:specter:had:drjj:75}) \drjj and (\ref{fig:specter:had:ptrat:25}, ~\ref{fig:specter:had:ptrat:75}) \ptrat distributions to samples with \frw$=0.25$ (left) and \frw$=0.75$ (right) using EMDs and sEMDs at hadronization level.}
 \label{fig:specter:had:2}
\end{figure}

Figures~\ref{fig:specter:had:1}--\ref{fig:specter:had:2} provide a comparison of the sEMD, EMD, and the Ref.~\cite{Andersen:2021mvw} reweightings. 
For \frw$=0.25$, the performance of the EMD and sEMD reweighting strategies is nearly identical.
For \frw$=0.75$, the differences are still small, though the sEMD shows slightly worse agreement for \Ht around 150 GeV, and slightly larger deviations for \njets $< 5$.
In both cases, the deviations are within the statistical uncertainties of the original sample. 

The Euclidean reweighting, labeled as `Andersen \emph{et al.},' shows a distinctively different behavior, particularly for \frw$=0.75$, where both the \Ht and \njets distributions have deviations well beyond statistical uncertainties.
Even with \frw$=0.25$, the \njets distribution shows worse agreement for \njets$\leq 2$, and slight tension for intermediate \Ht values.
We note that this metric was developed and studied in the context of much larger event samples than those utilized in this study: some aspects of its performance may be due to its application to a much sparser metric space.

\begin{figure}
    \centering
    \includegraphics[width=0.67\linewidth]{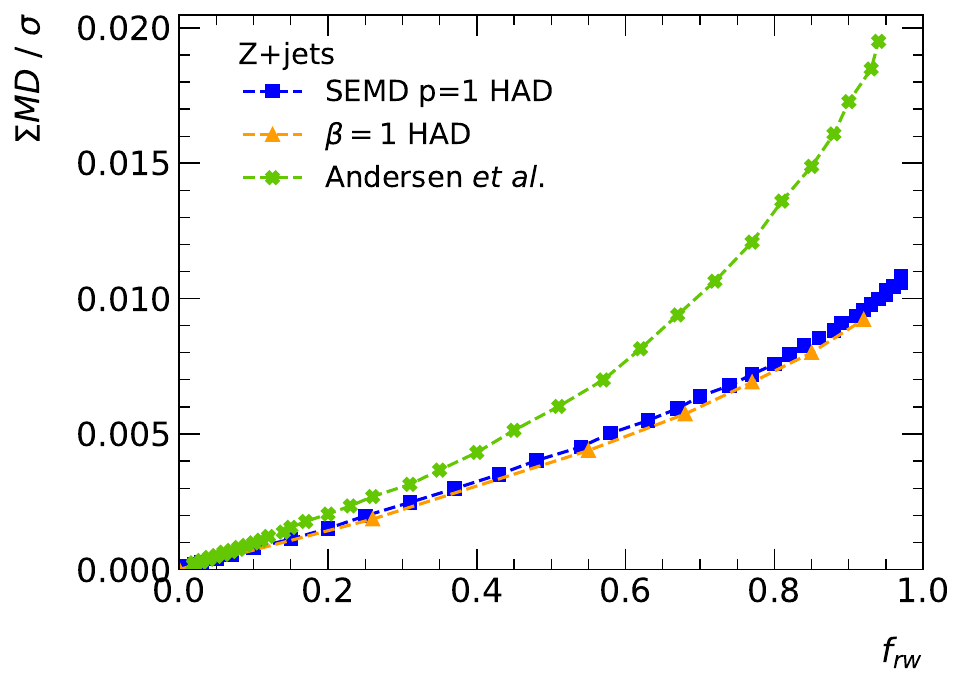}
    \caption{A comparison of the \XMD for the EMD, sEMD and the Ref.~\cite{Andersen:2021mvw} reweightings.}
    \label{fig:specter:xmd}
\end{figure}

Figure~\ref{fig:specter:xmd} shows the \XMD of these reweightings as a function of \frw, which is consistent with conclusions drawn from the 1D kinematic distributions. 
The sEMD shows similar but consistently worse performance than the EMD, while the Ref.~\cite{Andersen:2021mvw} reweighting has a larger \XMD across all values of \frw.
At high \frw values, the \XMD for the metric proposed in Ref.~\cite{Andersen:2021mvw} is nearly double that of the EMD.

%%%%%%%%%%%%%%%%%%%%%%%%%%%%%%%%%%%%%%%%%%%%%%%%%%%%%%%%%%%%%%%%%%%%%%%%%%%%%%%

\FloatBarrier
\subsection{Performance in \texorpdfstring{\ttbar}{ttbar} events}\label{sec:results:ttbar}

\begin{figure}[ht!]
    \centering
    \includegraphics[width=0.67\linewidth]{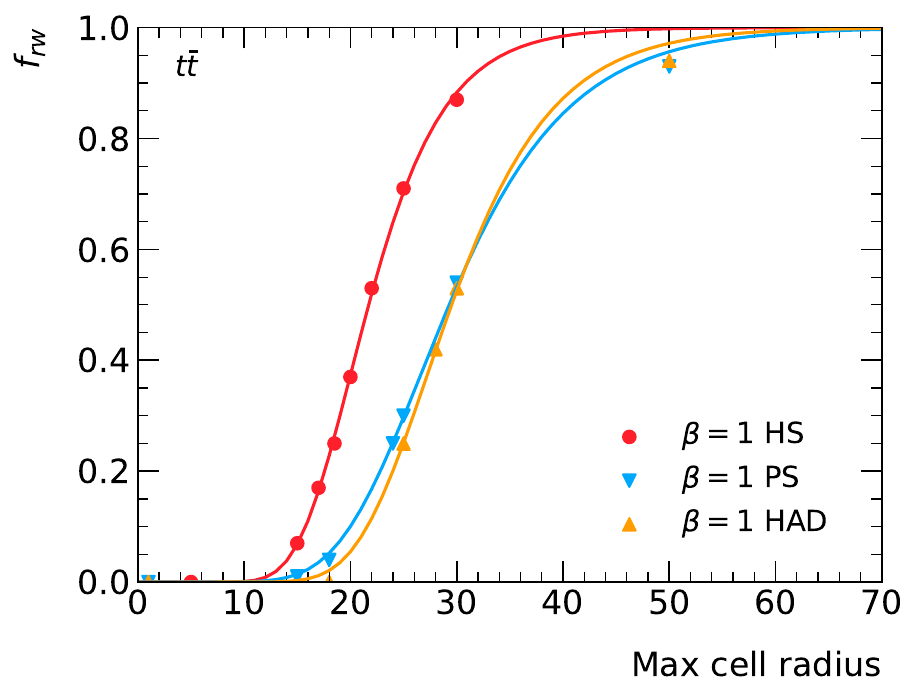}
\caption{The distribution of \frw as a function of the maximum cell radius for \ttbar events reweighted at matrix-element, parton shower, and hadronization level.
The distributions are fit using a Richards' curve~\cite{10.1093/jxb/10.2.290}.}
    \label{fig:ttbar:radii}
\end{figure}

Figure~\ref{fig:ttbar:radii} shows \frw as a function of the maximum cell radius in \ttbar events.
When compared to \zjets events (Figure~\ref{fig:stages:reweightfrac}), the maximum cell radius is larger in \ttbar events at a given \frw value.
With the increased complexity of the \ttbar events, a larger cell radius is necessary to achieve the same fractional reweighting.

\begin{figure}[htpb]
\centering{
   \subfloat[]{\includegraphics[width=0.44\textwidth]{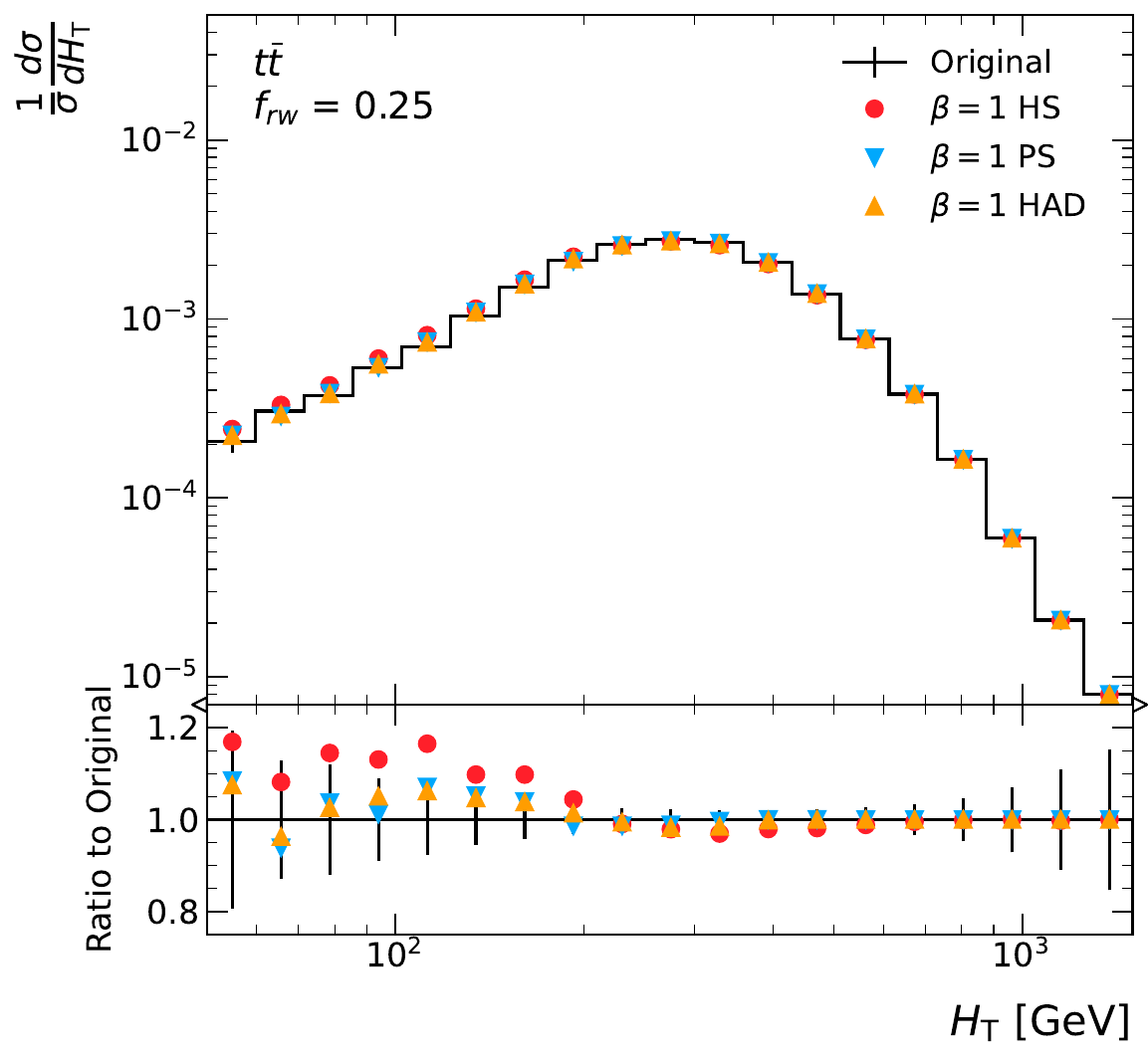}\label{fig:ttbar:stages:ht:25}}  
   \subfloat[]{\includegraphics[width=0.44\textwidth]{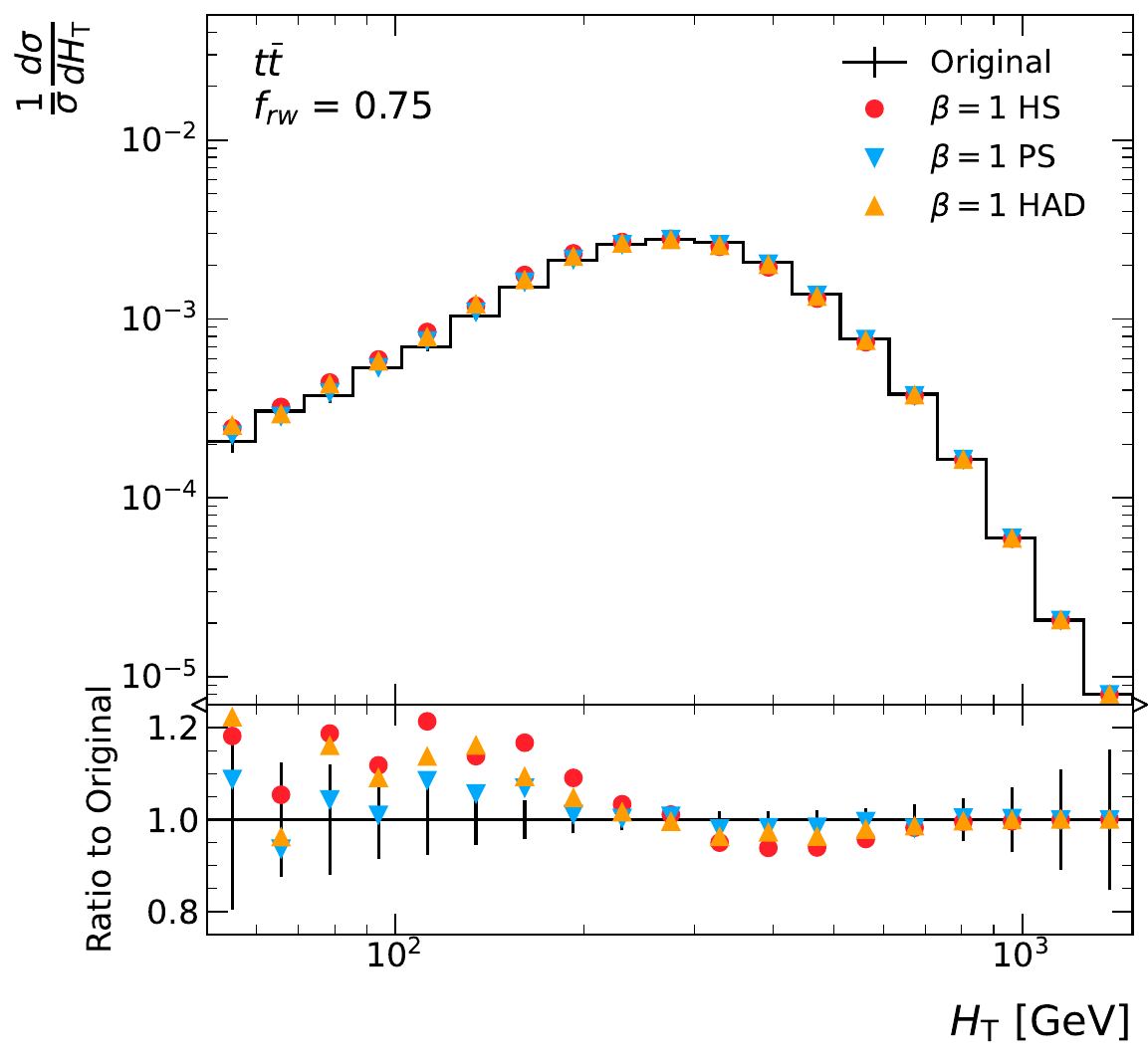} \label{fig:ttbar:stages:ht:75}}\\
   \subfloat[]{\includegraphics[width=0.44\textwidth]{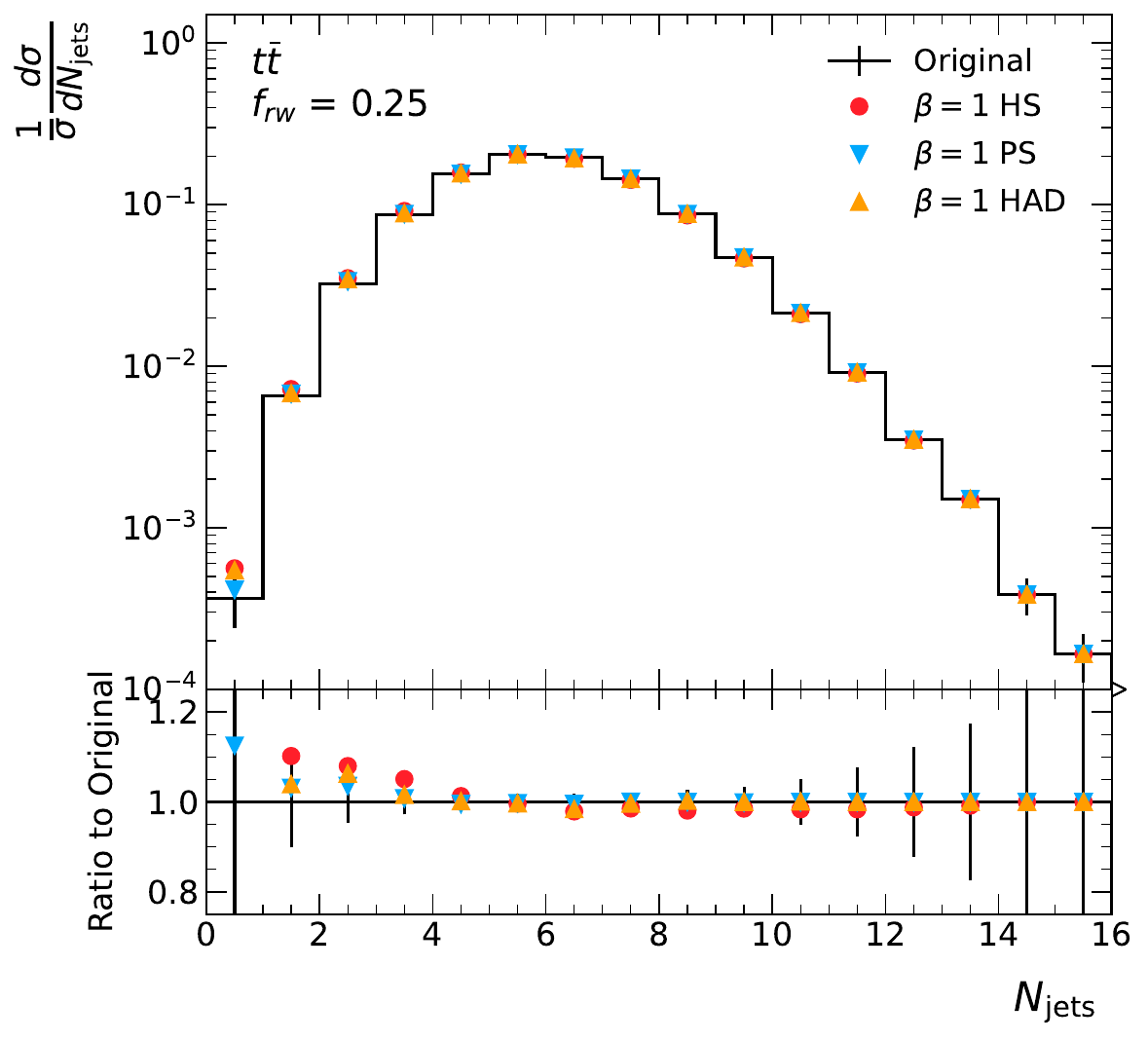} \label{fig:ttbar:stages:njets:25}}    
   \subfloat[]{\includegraphics[width=0.44\textwidth]{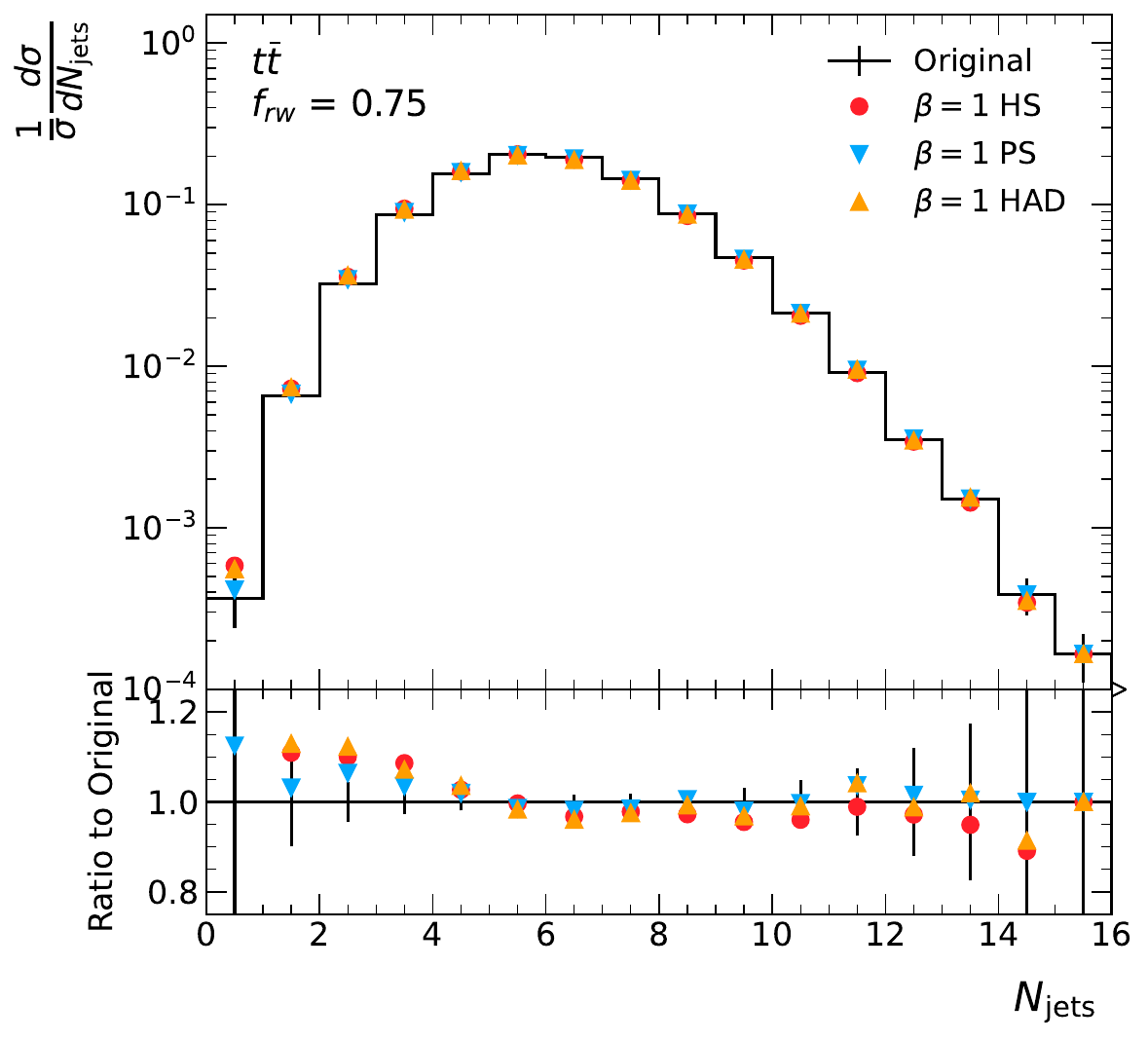} \label{fig:ttbar:stages:njets:75}}\\
   \subfloat[]{\includegraphics[width=0.44\textwidth]{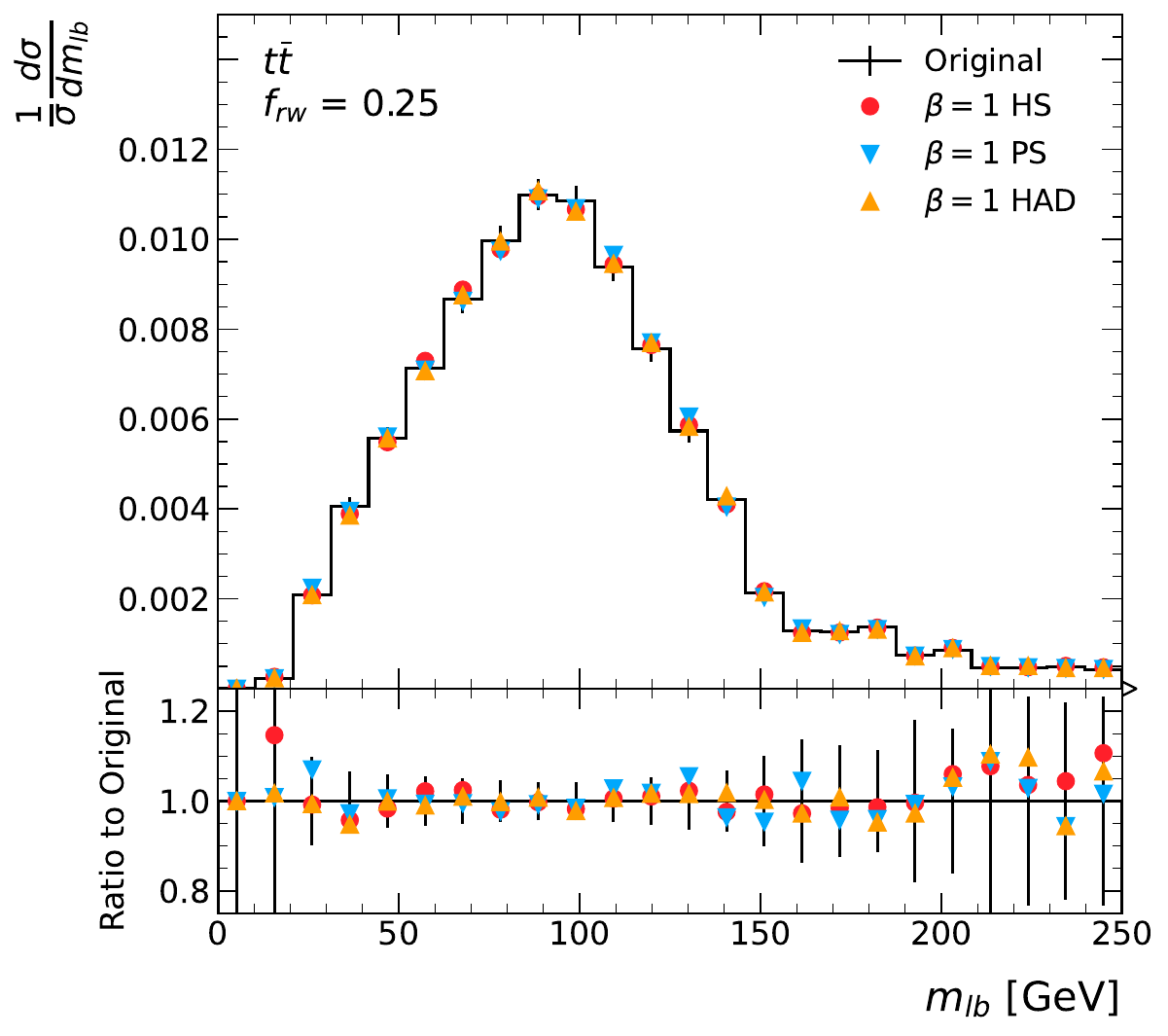} \label{fig:ttbar:stages:mlb:25}}
   \subfloat[]{\includegraphics[width=0.44\textwidth]{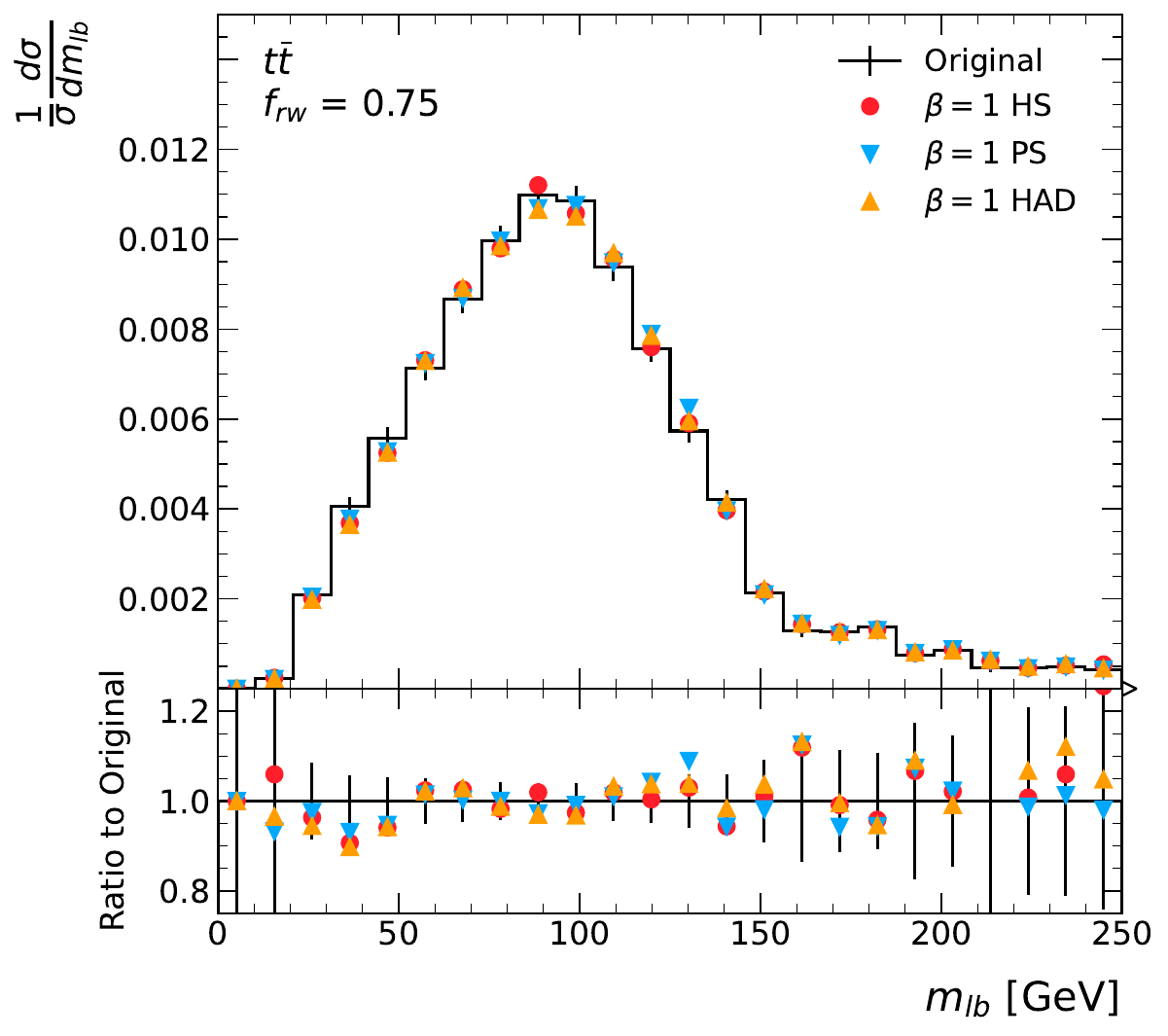} \label{fig:ttbar:stages:mlb:75}}
   \caption{Comparison of the unweighted \ttbar \Ht, \njets, and \mbl distributions to samples reweighted using EMDs at different generation stages, with \frw=0.25 (left) and 0.75 (right).}
   \label{fig:ttbar:stages}
   }
\end{figure}

Figure~\ref{fig:ttbar:stages} shows a comparison of kinematic distributions in \ttbar events for reweightings based on the different stages of event generation.
The HS reweighting is observed in tension with the original sample for both the \Ht and \njets distributions, even with the lower \frw$=0.25$. 
The PS and HAD reweightings reduce this disagreement, but still bias the results for \frw$=0.75$.
For both \frw$=0.25$ and \frw$=0.75$, the \mbl distributions have minimal bias, with the largest deviations observed for \mbl $<50$ GeV.

\begin{figure}[htbp]
    \centering
    \includegraphics[width=0.67\linewidth]{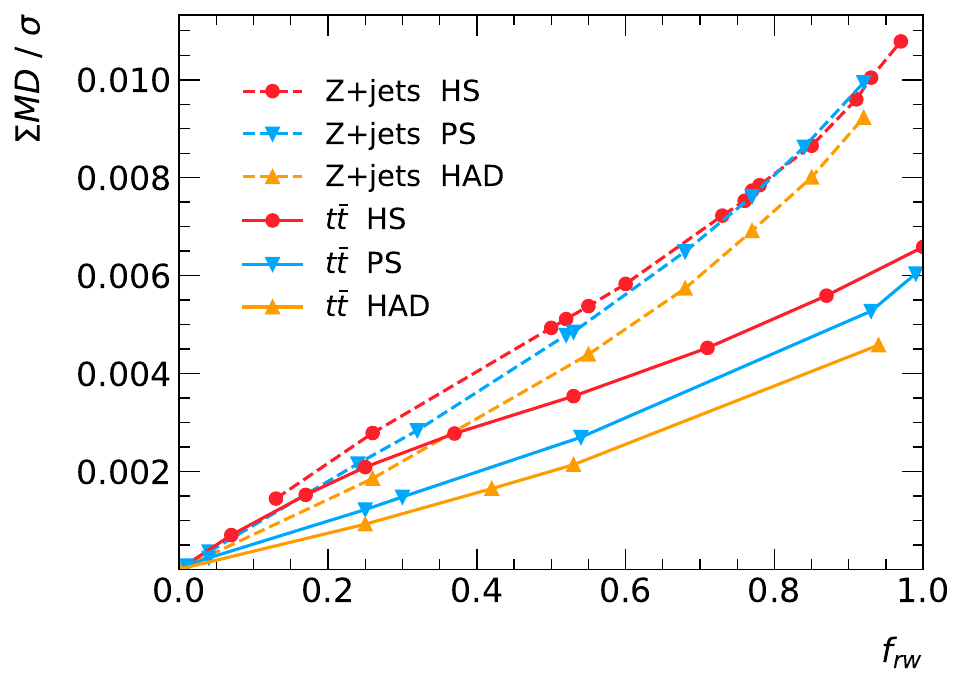}
    \caption{The \XMD, divided by the total cross section, for \ttbar and \zjets events.}
    \label{fig:ttbar:xmd}
\end{figure}

The \XMDs for \ttbar and \zjets events at different stages of generation are compared in Figure~\ref{fig:ttbar:xmd}.
Similar trends are seen for \ttbar events as for \zjets events, lending confidence that conclusions drawn based on studies of one physical process are applicable more broadly.
In general, the bias at a given \frw is smaller for \ttbar than for \zjets, likely due to the smaller initial fraction of negative weights in the \ttbar sample.
In both cases, the reweighting from hadronization shows the best performance.

%%%%%%%%%%%%%%%%%%%%%%%%%%%%%%%%%%%%%%%%%%%%%%%%%%%%%%%%%%%%%%%%%%%%

\FloatBarrier
\section{Concluding remarks}

In summary, these results establish optimal-transport-based cell resampling as a practical afterburner for reducing negative weights in collider MC simulations, directly addressing one of the computational pressures facing the High-Luminosity LHC era.
In this work, we have studied optimal-transport-based metrics, namely the EMD and its spectral variant the sEMD, as distance measures for cell resampling algorithms introduced in earlier particle physics literature by Andersen \emph{et al.}~\cite{Andersen:2021mvw}.
We have demonstrated that cell resampling algorithms using these metrics can be applied to mitigate negative and pathological event weights in Monte Carlo simulations of collider events, while incurring minimal bias in NLO \zjets and \ttbar samples.
Because these OT-based metrics are IRC-safe by construction, the resampling can be performed directly on final-state particles without the intermediate jet clustering stage required by previous Euclidean approaches and the attendant dependence this introduces on a jet definition and its parameters.
This IRC-safety also allows the reweighting to be applied after different stages of event generation: we studied reweightings based on Born-level hard-scattering information, after the parton shower and after hadronization.
Reweighting based on information from the fully-hadronized event was found to give the most stable performance across all observables and figures of merit considered, and consistent trends across \zjets and \ttbar samples lend confidence that these conclusions generalize across final-state topologies.
General-purpose algorithms like the cell resampling algorithm described in this work require no sample-specific training, and therefore could be applied in a complementary way with approaches that mitigate negative weights by learning a correction tailored to a specific MC sample~\cite{Nachman:2025lid,Palmer:2025jmb}.

Several quantitative conclusions follow from these studies.
After studying the performance of EMD reweighting with various choices of the angular parameter $\beta$, $\beta=1$ was found to give the best combination of accuracy and computational efficiency.
Choices of $\beta=0$ and $\beta\rightarrow\infty$ introduce substantial bias by neglecting the spatial distribution of energy within events.
When applied to the relatively small samples of events used in these studies, the OT-based metrics were found to outperform Euclidean object-based approaches introduced in earlier literature when compared in terms of the bias cell reweighting incurs on 1D kinematic distributions.

Finally, the Cross-Section Mover's Distance was introduced as a sample-level OT-based figure of merit that provides a single, holistic distance between two samples independent of any choice of observable or binning.
We have utilized a simple constant-offset procedure to extend the \XMD to samples containing negative weights, which is equivalent to a rigorous signed-measure formulation but requires no modification to a standard OT solver.
Though we have used it here to benchmark cell resampling, the \XMD is applicable broadly to any setting where potential biases in a full phase-space reweighting should be assessed in an unbinned manner.

% Python notebooks that reproduce these studies have been made available via GitHub~\cite{}.
%
% Compatible Monte Carlo simulated event samples that were used to perform these studies have been made available \emph{via} the CERN Zenodo platform~\cite{}.

\acknowledgments

We thank Jeppe Andersen and Andreas Maier for many interesting discussions on the topic of cell reweighting algorithms at L'\'Ecole de physique des Houches.
We thank Jeppe and Jesse Thaler for their prompt and helpful feedback on an early manuscript of this work.
We also wish to thank Rikab Gambhir for his technical support with the \textsc{Specter} package.

We are deeply grateful for the support this work obtained in its early stages in the form of a seed grant from the Brown University Data Science Institute.
This material is based on work supported by the U.S. Department of Energy, Office of Science, Office of High Energy Physics under Award Number DE-SC0026285.
This work is supported by the National Science Foundation under Cooperative Agreement PHY-2019786 (The NSF AI Institute for Artificial Intelligence and Fundamental Interactions, http://iaifi.org/).

\FloatBarrier
\bibliographystyle{JHEP}
\bibliography{ref,ATLAS,CMS,PubNotes,ConfNotes,ATLAS-useful}

\end{document}